\documentclass{aa}  

\usepackage{graphicx}
\usepackage{txfonts}
\usepackage{lipsum}
\usepackage{lscape}             
\usepackage{placeins}           

\usepackage{natbib}

\usepackage{comment}

\usepackage[colorlinks=true, citecolor=blue, linkcolor=blue, urlcolor=blue]{hyperref}

\def\apjl{Astrophys. J.}

\def\aap{Astron. Astrophys.}
\def\apjs{Astrophys. J. Suppl.}

\def \HI{{\sc Hi}}

\newcommand{\be}{\begin{equation}}
\newcommand{\e}{\end{equation}}
\newcommand{\bear}{\begin{eqnarray}}
\newcommand{\ear}{\end{eqnarray}}

\begin{document}


   \title{Modeling the \HI--Halo Connection: Evolution, Scatter, and a Halo-based Prescription for 21-cm Mock Catalogs}


%
%
%
\author{Mohd Kamran\inst{1,2}\corrauth{kamranmohd080@gmail.com}  
        \and Gabriella De Lucia\inst{1,2}
        \and Marta Spinelli\inst{3,4}
        \and Lizhi Xie\inst{5}  
        \and Fabio Fontanot\inst{1,2} 
        \and Michaela Hirschmann\inst{6}
        }
    
\institute{INAF -- Astronomical Observatory of Trieste, via G.B. Tiepolo 11, I-34131 Trieste, Italy
        \and IFPU -- Institute for Fundamental Physics of the Universe, via Beirut 2, I-34151 Trieste, Italy
        \and Observatoire de la C\^ote d'Azur, Laboratoire Lagrange, Bd de l'Observatoire, 06304 Nice, France
        \and Department of Physics \& Astronomy, University of the Western Cape, Cape Town 7535, South Africa
        \and Tianjin Astrophysics Center, Tianjin Normal University, Binshuixidao 393, Tianjin 300387, China
        \and Institute of Physics, Laboratory of Astrophysics, EPFL, Observatoire de Sauverny, Chemin Pegasi 51, 1290 Versoix, Switzerland
        }

\abstract{The redshifted 21-cm line of neutral hydrogen (\HI) is a powerful tracer of large-scale structure, and post-reionization \HI\ intensity mapping is emerging as a competitive cosmological probe whose interpretation requires a description of how \HI\ populates galaxies and dark matter halos. We characterize the \HI--halo mass relation, its redshift evolution, and its intrinsic scatter, identifying its secondary dependences. We use the updated GAlaxy Evolution and Assembly (GAEA) semi-analytic model, applied to the Millennium-I and Millennium-II simulations, to predict the \HI\ mass function (HIMF) and the \HI--halo mass relation from the present day to redshift $z\simeq5$. At $z=0$, the model reproduces the observed HIMF and its decomposition by host-halo mass. The median \HI--halo mass relation rises with halo mass, peaks near $10^{11.7}\,M_\odot$, declines as central galaxies are quenched by feedback from active galactic nuclei, and rises again where satellites dominate, approaching a single power law at high redshift. We show that the substantial scatter, of about 0.5 dex, is not random but is governed by halo assembly: at fixed mass, higher-spin, later-forming, and less-concentrated halos are systematically \HI-richer, with spin together with either concentration or formation time accounting for part of this scatter and leaving an intrinsic dispersion of about 0.3 dex. We encode the median relation, these secondary trends, and the intrinsic scatter in a compact, physically motivated prescription expressed entirely in terms of quantities available in dark-matter halo catalogs. This prescription reproduces the full scatter and enables the construction of large-volume 21-cm mock catalogs for interpreting ongoing intensity-mapping measurements with SKA precursor facilities, such as MeerKAT, and for preparing for forthcoming surveys with the SKA.}


\titlerunning{Modeling the \HI--halo connection}
\authorrunning{Kamran et al.}

\maketitle

\nolinenumbers

\section{Introduction}
\label{sec:intro}
Constraining the large-scale structure (LSS) of the Universe is a central goal of modern cosmology, because the matter distribution encodes both the cosmic expansion history and the growth of density perturbations, thereby providing stringent tests of the $\Lambda$CDM paradigm and its possible extensions. Neutral atomic hydrogen (\HI) offers a physically well-motivated tracer of this structure: it is detectable in emission through the redshifted 21-cm hyperfine transition and, on sufficiently large scales, traces the underlying matter distribution, albeit as a biased tracer whose small-scale content is regulated by galaxy-formation processes \citep{furlanetto06,pritchard12}. Once reionization is complete, this signal originates predominantly from dense, self-shielded systems that are largely associated with galaxies. Because the observed frequency maps directly onto redshift, the 21-cm line enables three-dimensional tomographic studies of LSS: radio observations between $1420$ and $\sim200$ MHz access the signal from $z=0$ out to redshifts approaching the end of reionization ($z\sim6$) \citep{Wyithe:2008mv,bull15}. In this post-reionization regime, the 21-cm signal is intimately linked to galaxy-formation physics, yet it can simultaneously be exploited as a cosmological observable through its connection to the underlying matter field. At the redshifts relevant for cosmological surveys, however, the galaxies hosting this \HI\ are generally too faint to be detected individually in the numbers required for precision clustering over cosmological volumes.

This limitation is circumvented by 21-cm \emph{intensity mapping}, which measures the integrated line emission in large three-dimensional voxels and uses the unresolved brightness-temperature fluctuations to map the underlying large-scale structure \citep{Bharadwaj:2000av,Bharadwaj:2003uh,bharadwaj04b,Chang:2007xk,Loeb:2008hg,bull15}. The Square Kilometre Array (SKA) will transform \HI\ astronomy: it will substantially expand direct \HI\ surveys of individual galaxies at low redshift and, through intensity mapping, extend clustering measurements to the higher redshifts and larger volumes where complete individual-galaxy samples are impractical \citep{SKA_SWG20}. Intensity mapping is therefore particularly well suited to late-time cosmology, where it can access linear and mildly non-linear scales over broad redshift ranges, thereby constraining the expansion history and the growth rate of structure \citep{Chang:2007xk,Wyithe:2007rq,Seo:2009fq,Masui:2010mp,bull15,obuljen18}. Its cosmological interpretation, however, requires more than a geometrical mapping of LSS: it depends on the \HI\ population, including its abundance and distribution across galaxies and dark matter halos, and hence on how it traces the underlying matter field. These properties are shaped by galaxy formation and assembly, and must be modeled and constrained by observations.

The most direct constraints on the \HI\ population come from low-redshift 21-cm surveys that resolve individual galaxies. Blind surveys, notably the \HI\ Parkes All-Sky Survey (HIPASS; \citealt{zwaan05}) and the Arecibo Legacy Fast ALFA Survey (ALFALFA; \citealt{martin10,jones18}), have measured the local \HI\ mass function (HIMF) and the corresponding cosmic \HI\ density parameter, $\Omega_{\rm HI}$, and have constrained the dependence of \HI\ content on galaxy properties and environment. These measurements provide a low-redshift benchmark for the models used to interpret intensity-mapping data.

Beyond the local Universe, the \HI\ content of individual galaxies has been probed to intermediate redshift through 21-cm spectral stacking, in which the emission from many galaxies with known optical redshifts is co-added to recover an average \HI\ mass below the individual-detection threshold. Giant Metrewave Radio Telescope (GMRT) observations have been used to constrain the mean \HI\ mass of star-forming galaxies at $z=0.24$ and $z=0.37$ \citep{lah07,lah09}, and the technique has since been extended to $z\approx1.3$ \citep{kanekar16}. Stacking therefore bridges resolved surveys in the local Universe and intensity-mapping measurements at higher redshift, but it yields the average \HI\ content of selected galaxy samples rather than the full distribution of \HI\ masses required to determine the HIMF.

At comparable redshifts, the post-reionization 21-cm signal has been detected statistically through intensity mapping, most robustly in cross-correlation with optically selected large-scale-structure tracers. Radio foregrounds and many instrumental systematics are uncorrelated with the external galaxy field, making this approach comparatively insensitive to them, although these contaminants can still increase the measurement uncertainty. Green Bank Telescope (GBT) data have been cross-correlated with DEEP2 \citep{chang10} and with WiggleZ \citep{masui13}, and Parkes data with the 2dF galaxy survey \citep{anderson18}. More recently, the MeerKLASS Collaboration \citep{meerklass25} reported a $>4\sigma$ detection of the cross-power spectrum between MeerKAT single-dish intensity maps and GAMA galaxies at $0.39<z<0.46$, on scales $k<0.3\,h\,{\rm Mpc}^{-1}$. Auto-power-spectrum measurements are more demanding because foreground emission---dominated by Galactic synchrotron radiation---and residual instrumental systematics contribute directly to the radio auto-spectrum \citep{switzer13}. Foreground characterization, removal, and control of signal loss are therefore essential for both auto- and cross-correlation analyses \citep{ghosh12,carucci25,spinelli26}. Direct detections of the 21-cm \HI\ auto-power spectrum have now been reported with MeerKAT interferometric data at $z\approx0.32$ and $0.44$, on Mpc scales \citep{paul26}, and with CHIME at $1.01<z<1.34$, over $0.4\,h\,{\rm Mpc}^{-1}\lesssim k\lesssim1.5\,h\,{\rm Mpc}^{-1}$, with a detection significance of $12.4\sigma$ \citep{chime25}. Extending auto-power-spectrum measurements to larger cosmological scales and higher redshifts is a key goal of forthcoming surveys.. 

Interpreting these measurements requires a physical description of the \HI\ population and of its connection to the matter distribution. On large scales, the two-point statistics of the post-reionization 21-cm signal are commonly characterized by the mean brightness temperature, which is set by $\Omega_{\rm HI}$ at fixed cosmology, the effective \HI\ bias, $b_{\rm HI}$, and a shot-noise contribution arising from the discrete distribution of \HI-bearing systems \citep{Wyithe:2008mv,castorina17,obuljen18,spinelli20}. This large-scale description is only one consequence of the \HI--galaxy--halo connection. The HIMF, its environmental dependence, nonlinear clustering, and the construction of realistic mock observations all depend on how \HI\ is distributed across galaxy and halo populations and on how that distribution evolves with redshift. Several approaches have been developed to model these aspects. Empirical prescriptions, including halo-model parameterizations and abundance-matching approaches, specify the relation between \HI\ and halo mass using observational constraints \citep{bagla10,padmanabhan17,obuljen18}; they are computationally inexpensive and can be applied to arbitrarily large volumes, but predictions outside the range of the constraining data remain uncertain. Cosmological hydrodynamical simulations follow gas dynamics and a range of baryonic processes directly, but their computational cost limits the volume and resolution that can be achieved simultaneously, and the neutral and molecular gas fractions are often assigned in post-processing \citep{duffy12,dave13,lagos15,Villaescusa-Navarro:2018vsg}. Semi-analytic models (SAMs) such as \textsc{shark} \citep{lagos18} and GAEA \citep{xie17,xie20} offer an intermediate route: by coupling parameterized baryonic physics to large dark-matter merger trees, they can be applied to large volumes while retaining the assembly histories, environments, and central--satellite structure of their halos. This combination is well suited to cosmological applications, which require large volumes, ideally sampled by multiple independent lightcones, together with a physically informed \HI\ population, and SAMs have accordingly been used extensively to investigate the \HI\ content of galaxies and halos \citep{kim17,baugh19,chauhan20,spinelli20}. To predict 21-cm observables, the modeled \HI\ distribution is converted into a redshift-space brightness-temperature field using the positions and velocities of \HI-bearing galaxies, which provides the basis for constructing mock maps \citep{Villaescusa-Navarro:2018vsg,spinelli20}.

In this work, we use the semi-analytic model GAlaxy Evolution and Assembly (GAEA) to characterize the \HI\ population of galaxies and dark matter halos across cosmic time, providing the inputs needed to model and interpret post-reionization 21-cm observations. We examine the HIMF and its dependence on host-halo mass, the \HI--halo mass relation, $M_{\rm HI}(M_{\rm h})$, its intrinsic scatter, and the secondary halo properties that contribute to this scatter. This work extends the analysis of \citet{spinelli20}, which was based on an earlier version of the GAEA model (GAEA2017). Here we employ the updated GAEA2023 model \citep{delucia24b}, which incorporates revised treatments of satellite gas stripping and AGN feedback, and compare our results with an independent SAM \citep{chauhan20} and with available observational constraints. Building on these results, we derive a compact prescription for $M_{\rm HI}(M_{\rm h})$ and its scatter, expressed entirely in terms of properties available in dark-matter-only simulations, enabling large volumes to be populated with \HI\ for the construction of 21-cm mock catalogs.

The structure of this paper is as follows. In Sect.~\ref{sec:simulation}, we describe the GAEA2023 semi-analytic model and the simulation data used in this work. In Sect.~\ref{sec:himf}, we present the HIMF and conditional HIMF, and examine their evolution with redshift. In Sect.~\ref{sec:Mh_MHI}, we investigate the relation between the total \HI\ content of dark matter halos and halo mass, together with its evolution. In Sect.~\ref{sec:physical drivers}, we interpret the physical origin of the \HI--halo mass relation by exploring its dependence on secondary halo properties. In Sect.~\ref{sec:hihm_fit}, we introduce analytic fitting functions for the median relation and its scatter. Finally, Sect.~\ref{sec:conclusions} summarizes our main results and discusses their implications for the \HI\ prescriptions needed to interpret ongoing 21-cm intensity-mapping measurements from SKA precursor facilities and to prepare for forthcoming SKA surveys.

\section{Modeling \HI\ in Galaxies and Dark Matter Halos}
\label{sec:simulation}

\subsection{Numerical Simulations}

This work is based on galaxy catalogs generated with the semi-analytic model GAEA\footnote{https://sites.google.com/inaf.it/gaea/}. The GAEA model has been applied to merger trees extracted from two large cosmological dark matter simulations: the Millennium I (MSI; \citealt{springel05}) and Millennium II (MSII; \citealt{boylan09}) runs. While MSI provides a larger simulation volume, MSII offers superior mass resolution. Both simulations adopt a WMAP1 cosmology \citep{spergel03} with parameters $\Omega_\mathrm{m}=0.25$, $\Omega_\Lambda=0.75$, $\Omega_\mathrm{b}=0.045$, $h=0.73$, and $\sigma_8=0.9$. The numerical parameters of these simulations are summarized in Table~\ref{tab:table1}. Throughout this work, a Chabrier initial mass function is assumed \citep{chabrier03}.

Initial conditions for both the MSI \citep{springel05} and MSII \citep{boylan09} simulations were established at $z=127$, and both simulations were run using different versions of the \textsc{gadget} $N$-body code \citep{springel01b,springel05}. Halos and subhalos were identified on-the-fly using a friends-of-friends (FoF) algorithm \citep{davis85} with linking length $b=0.2$, followed by the \textsc{subfind} substructure finder \citep{springel01a}, retaining only subhalos containing at least 20 bound particles. Halo and subhalo catalogs were stored at 64 discrete output times (``snapshots'') for MSI and at 68 for MSII. Subhalo merger trees suitable for running GAEA were constructed following the methodology described in \citet{springel05}. Within each FoF group, the galaxy hosted by the main (most-bound) subhalo is the central (\textit{type~0}), and galaxies hosted by other resolved subhalos are satellites (\textit{type~1}). When a satellite's subhalo is tidally stripped below the resolution limit, its galaxy is retained as an ``orphan'' (\textit{type~2}), its position followed via the most-bound particle of the disrupted subhalo until it merges with the central on a residual dynamical-friction timescale \citep{delucia07,delucia10}. Throughout the following sections, ``satellites'' denotes \textit{type~1} $+$ \textit{type~2} galaxies.

\subsection{Semi-analytic model: GAEA}

The GAEA model employed here builds upon the original framework of \citet{delucia07} and incorporates a series of targeted improvements developed over successive years. These include: (a) a detailed treatment of non-instantaneous recycling of gas, metals, and energy that accounts for stellar lifetimes and enables tracking of individual metal species \citep{delucia14}; (b) an enhanced stellar feedback scheme calibrated against high-resolution hydrodynamical simulations to match the galaxy stellar mass function to $z\sim 3$ \citep{hirschmann16}; (c) explicit partitioning of cold gas into atomic (\HI) and molecular (${\rm H_2}$) phases, coupled to molecular-based star formation \citep{xie17}; (d) refined angular momentum evolution for gas and stellar discs, together with gradual ram-pressure stripping of satellite gas reservoirs \citep{xie20}; and (e) improved modeling of cold gas accretion onto supermassive black holes and associated AGN-driven outflows \citep{fontanot20}. This configuration has been calibrated to reproduce the galaxy stellar mass function to $z<3$, local \HI\ and ${\rm H_2}$ mass functions, and the evolution of the AGN luminosity function to $z<4$. This work employs the upgraded GAEA model presented in \citet[designated therein as GAEA2023]{delucia24b}, which integrates the advanced treatments of satellite gas stripping and AGN physics described in points (d) and (e) above within a unified framework that also incorporates all preceding developments. 

\begin{table}
\caption{Main numerical parameters of the simulations used in this study. From left to right: comoving box size $L_\mathrm{box}$ $[h^{-1}\,\mathrm{Mpc}]$, number of particles $N_\mathrm{p}$, dark matter particle mass $M_\mathrm{p}$ $[h^{-1}\,\mathrm{M}_\odot]$, minimum resolved stellar mass $M_{\mathrm{s,min}}$ $[h^{-1}\,\mathrm{M}_\odot]$, and minimum resolved dark matter subhalo mass $M_{\mathrm{h,min}}$ $[h^{-1}\,\mathrm{M}_\odot]$ considered in the analysis.}
\label{tab:table1}
\centering
\setlength{\tabcolsep}{4.9pt}
\begin{tabular}{lccccc}
\hline\hline
Simulation & $L_{\rm box}$ & $N_{\rm p}$ & $M_{\rm p}$ & $M_{\rm s,min}$ & $M_{\rm h,min}$ \\
\hline
MSI  & 500 & $2160^3$ & $8.6 \times 10^8$ & $10^8$ & $1.7 \times 10^{10}$ \\
MSII & 100 & $2160^3$ & $6.8 \times 10^6$ & $10^6$ & $1.4 \times 10^8$ \\
\hline
\end{tabular}
\end{table}

\section{The galaxy \HI\ mass function (HIMF)}
\label{sec:himf}

\begin{figure}
	\includegraphics[width=\columnwidth]{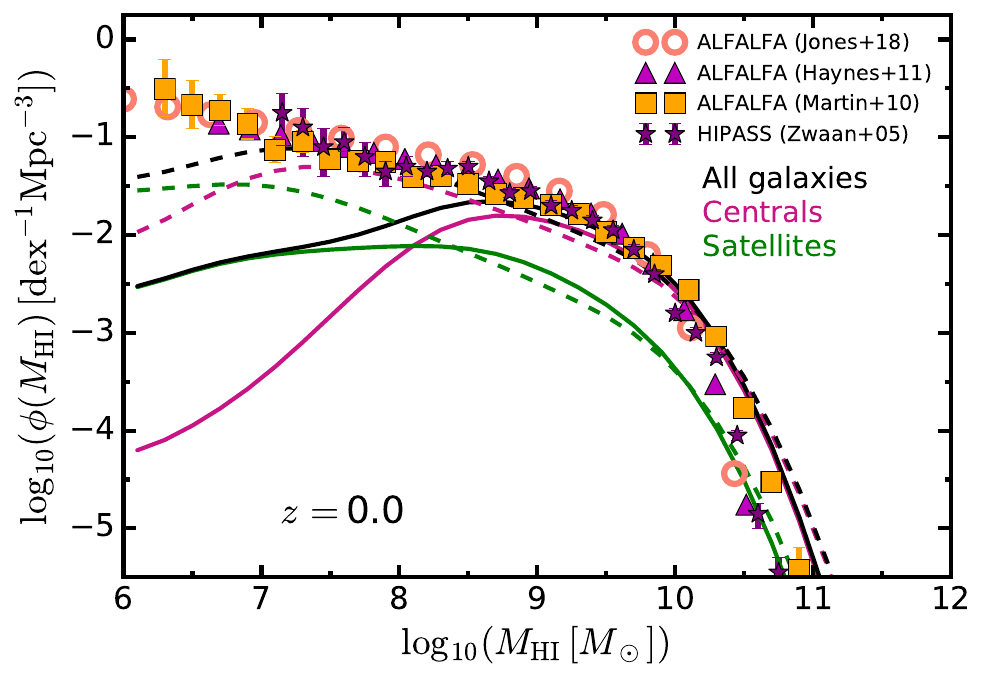}
    \caption{The galaxy \HI\ mass function (HIMF) at $z=0$ from the MSI (solid curves) and MSII (dashed curves) simulations. Black curves show the HIMF of all galaxies (centrals and satellites combined); magenta and green curves show the central- and satellite-galaxy contributions, respectively. Observational data points from HIPASS \citep{zwaan05} and ALFALFA \citep{martin10,haynes11,jones18} are shown for comparison.}
    \label{fig:himf_z0}
\end{figure}

In this section, we present the galaxy HIMF, i.e. the comoving number density of galaxies per logarithmic interval in \HI\ mass, in the MSI and MSII simulations, separating the contributions of central and satellite galaxies. In GAEA2023, the partitioning of the cold gas into its atomic and molecular components is tuned so that the model reproduces the local ($z=0$) HIMF against the ALFALFA measurement of \citet{haynes11}. The resulting agreement at the high-mass end, above the blind-survey completeness limit ($\log(M_{\rm HI}/M_\odot)\gtrsim9$), is therefore partly by construction, whereas the low-mass end is less directly constrained by the calibration and therefore offers a more independent test of the model. To define a clean galaxy sample, we apply stellar-mass cuts following \citet{spinelli20}, retaining only galaxies with $M_{\ast}>10^6\,M_\odot$ in MSII (this limit corresponds to the stellar mass down to which the predicted stellar mass function matches the observed one), while for MSI a $M_{\ast}\sim10^8\,M_\odot$ cut is set by convergence with the higher-resolution MSII run. The adopted thresholds are listed in Table~\ref{tab:table1}. These are lower than the more conservative cuts ($M_{\ast}\sim10^9\,M_\odot$ for MSI and $\sim10^8\,M_\odot$ for MSII) used in earlier GAEA studies \citep{xie17,zoldan17}, and allow us to probe the low-\HI\ end of the HIMF, which is relevant for intensity mapping and faint-galaxy statistics. The faint-end results should nonetheless be interpreted with caution, given resolution limits and the growing impact of numerical and modeling uncertainties at low \HI\ masses.

\subsection{The HIMF at \texorpdfstring{$z=0$}{z=0}}
The HIMF at $z=0$ is shown in Fig.~\ref{fig:himf_z0}, comparing MSI and MSII runs with local measurements from HIPASS and ALFALFA. The model HIMFs are convolved with a Gaussian of width $0.25$~dex, an estimate of the observational uncertainty on \HI-mass estimates. We note that this convolution is applied only for the comparison with the data, while the model was calibrated on the unconvolved HIMF and the convolution is not applied to any of the intrinsic model quantities discussed in later sections. Over the common converged range, $9\lesssim\log(M_{\rm HI}/M_\odot)\lesssim11$, the MSI and MSII predictions agree to within $0.2$--$0.3$~dex. The larger volume of MSI provides better statistics for the rare systems at the high-\HI-mass end, whereas the higher mass resolution of MSII extends the reliable range down to $\log(M_{\rm HI}/M_\odot)\sim7$. The HIMF of all galaxies, comprising central and satellite galaxies, reproduces the HIPASS and ALFALFA determinations above their completeness limit ($\log(M_{\rm HI}/M_\odot)\gtrsim9$). At the massive end ($M_{\rm HI}\gtrsim10^{10}\,M_\odot$) the HIMF is dominated by central galaxies, whereas satellites contribute substantially at intermediate masses, consistent with the environmental gas removal that depletes satellite \HI\ reservoirs in GAEA (Sect.~\ref{sec:simulation}); the steeper decline of the central contribution below $\log(M_{\rm HI}/M_\odot)\sim9$ (MSI) and $\sim7$ (MSII) follows the differing resolution limits of the two runs rather than a physical feature of the model. A decomposition of the local HIMF by galaxy color is presented in Appendix~\ref{sec:himf_color}.

\begin{figure}
	\includegraphics[width=\columnwidth]{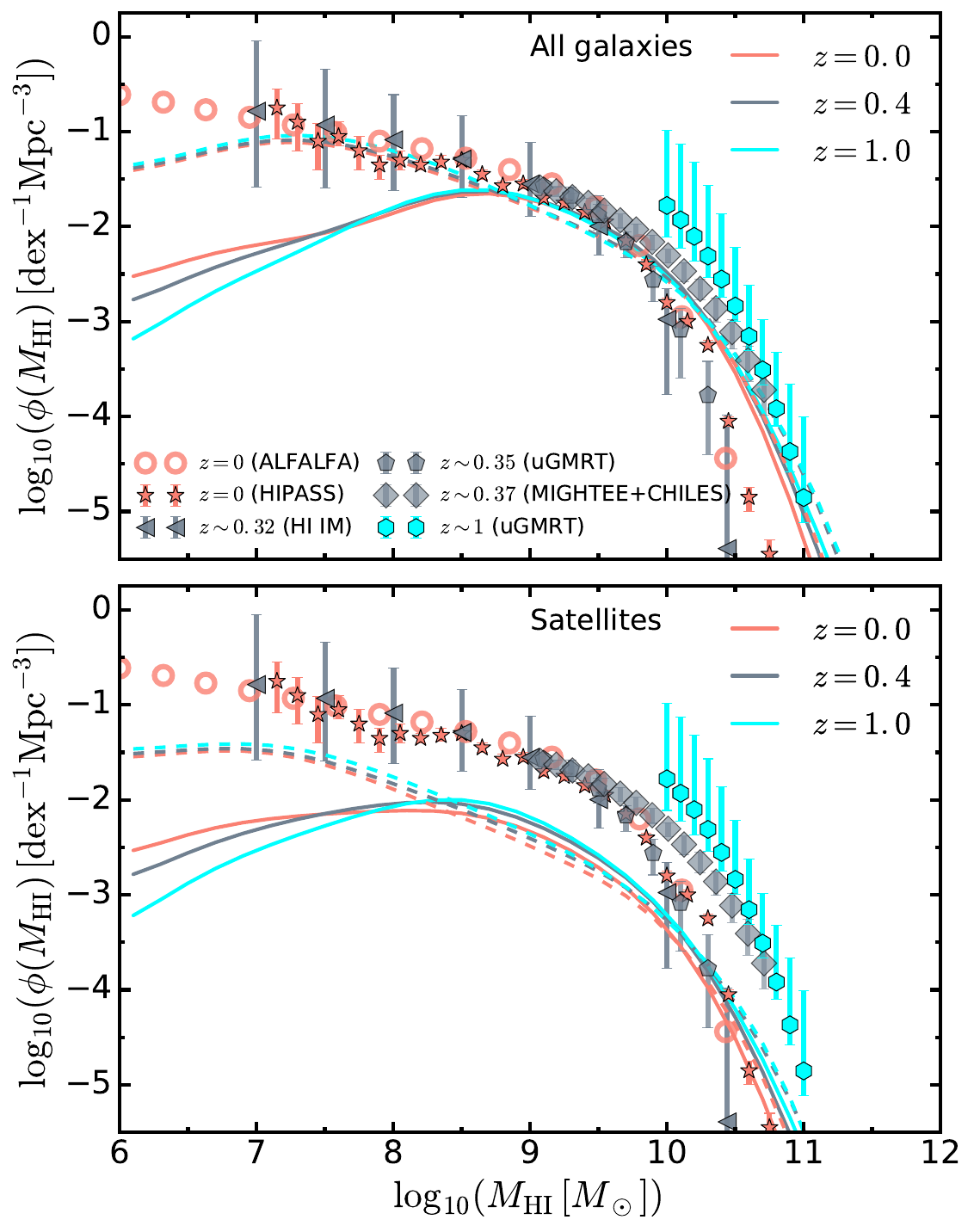}
    \caption{Evolution of the galaxy HIMF from $z=0$ to $z=1$. {\it Upper panel}: Total HIMF for all galaxies from MSI (solid) and MSII (dashed) simulations. {\it Lower panel}: Satellite-only HIMF from the same simulations. Colored lines correspond to different redshifts as indicated, with matching symbols showing observational constraints: at $z=0$, ALFALFA blind survey shown by salmon annuli \citep{jones18} and HIPASS blind survey shown by salmon stars \citep{zwaan05}; at $z\sim0.32$, slate gray triangles (MeerKAT intensity mapping; \citealt{paul26}); at $z\sim0.35$, slate gray pentagons (uGMRT stacking; \citealt{bera22}); at $z\sim0.37$, slate gray diamonds (MIGHTEE+CHILES stacking; \citealt{sinigaglia25}); at $z\sim1$, cyan hexagons (uGMRT stacking; \citealt{chowdhury24}). Model predictions have been convolved with a representative $0.25$\,dex observational uncertainty in \HI-mass determination.}
    \label{fig:himf_diff_z}
\end{figure}

The ALFALFA determinations differ slightly among themselves: the measurements of \citet{martin10} and \citet{haynes11} are based on the 40\% catalog ($\alpha.40$), whereas the final-catalog HIMF of \citet{jones18} has a flatter low-mass slope ($\alpha=-1.25$ versus $-1.33$) and a marginally lower ``knee'' mass, differences that \citet{jones18} attribute to sample variance and the differing large-scale structures sampled by the ALFALFA fields. Since GAEA2023 is calibrated on the earlier ALFALFA HIMF \citep{haynes11}, it follows those determinations most closely; relative to \citet{jones18}, adopted as the $z=0$ reference in Fig.~\ref{fig:himf_diff_z}, it therefore slightly overpredicts the abundance of the most \HI-massive systems.

\subsection{The HIMF for different \texorpdfstring{$z$}{z}}
The redshift evolution of the HIMF is presented in Fig.~\ref{fig:himf_diff_z}. The all-galaxy HIMF evolves only weakly between $z=0$ and $z\simeq1$, with the most noticeable evolution occurring at the high-mass end, while the knee and intermediate-mass regime remain nearly unchanged. This weak evolution up to $z\sim1$ is qualitatively consistent with previous semi-analytic predictions \citep{lagos11,baugh19,spinelli20}. At intermediate redshift ($z\sim0.3$--$0.4$), the predictions are broadly consistent with available indirect estimates, including the HIMF inferred from MeerKAT intensity mapping \citep{paul26} and the uGMRT \citep{bera22} and MIGHTEE+CHILES \citep{sinigaglia25} stacking-based estimates for star-forming galaxies; around $M_{\rm HI}\simeq10^{10}\,M_\odot$ the model at $z=0.4$ lies slightly below the \citet{sinigaglia25} estimate. At $z\sim1$, the star-forming-galaxy HIMF of \citet{chowdhury24} likewise lies above the model at the high-mass end. Because these stacking-based estimates refer to selected star-forming samples whereas the model curves include all galaxies, the comparisons are indicative rather than strictly like-for-like. Direct constraints on the total HIMF at $z\gtrsim0.3$ remain sparse and are affected by cosmic variance and stacking systematics \citep{bera22,sinigaglia25,chowdhury24}.

The lower panel shows that satellites contribute a sub-dominant fraction of the total HIMF at all three redshifts shown ($z=0$, $0.4$, and $1$). Notably, MSI and MSII converge well for the satellite HIMF above their respective resolution limits across this redshift range, indicating that this behavior is robust to numerical resolution.

\begin{figure}
	\includegraphics[width=\columnwidth]{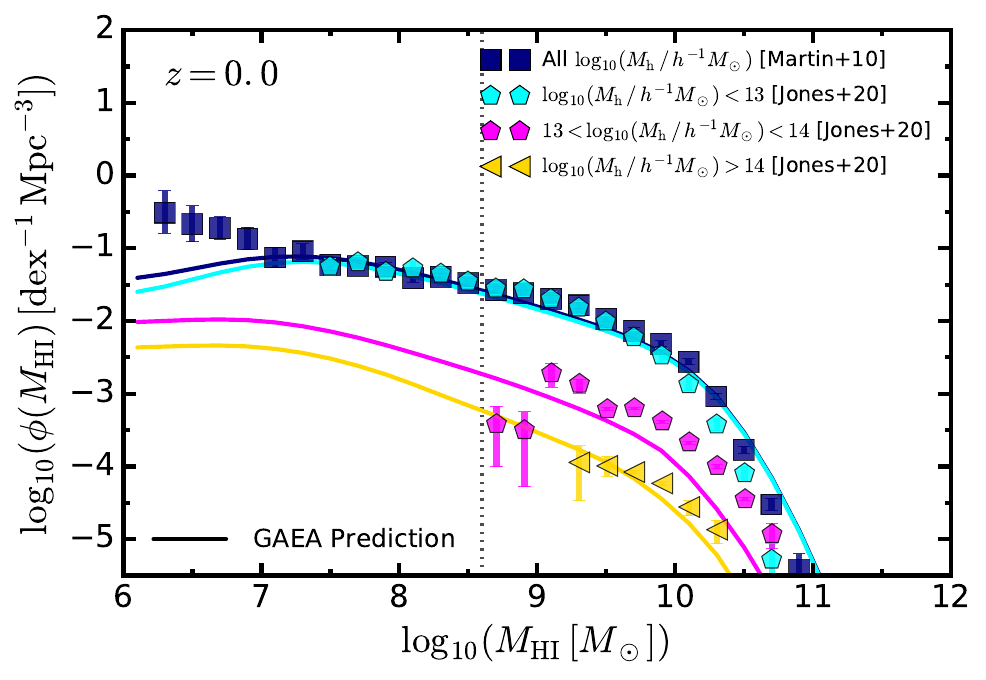}
    \caption{\HI\ conditional mass function at $z=0$ for different ranges of host halo mass. Solid curves show the predictions of the GAEA2023 model for all halos and for the three halo-mass bins indicated in the legend. Navy squares show the total ALFALFA HIMF of \citet{martin10} (i.e.\ the all-halos case), while the cyan, magenta, and gold symbols show the ALFALFA group-catalog measurements of \citet{jones20} in restricted host-halo mass ranges. Each model curve combines the MSI prediction at the high-mass end with the MSII prediction at the low-mass end, blended smoothly across the \HI\ mass at which the two simulations converge; the vertical dotted line marks this convergence scale, $\log(M_{\rm HI}/M_\odot)\simeq8.6$, below which the MSI prediction alone would turn over owing to its coarser mass resolution.}
    \label{fig:hicmf_z0_one_panel}
\end{figure}

\subsection{The \HI\ conditional mass function at \texorpdfstring{$z=0$}{z=0}}
The \HI\ conditional mass function at $z=0$ is shown in Fig.~\ref{fig:hicmf_z0_one_panel} for different host-halo mass ranges, comparing the GAEA2023 predictions with ALFALFA-based constraints. Here the conditional HIMF is defined per unit comoving volume: each halo-mass bin gives the contribution to the total HIMF from galaxies hosted by halos in that mass range, rather than a per-halo occupation function; by construction, the all-halos case coincides with the total HIMF. For all halos combined, the model reproduces the shape and normalization of the total HIMF of \citet{martin10} in the converged range ($\log(M_{\rm HI}/M_\odot)\gtrsim8.6$), where the measurements are well constrained. When the sample is split by host-halo mass, the model follows the trends measured in the ALFALFA group catalog of \citet{jones20}: the conditional HIMF shifts to progressively lower amplitudes for more massive halos. Because this per-volume definition retains the abundance of the host halos, this ordering reflects primarily the steeply declining space density of massive halos -- groups and clusters being far rarer than galaxy-scale halos -- with environmental processes additionally suppressing the \HI\ content of the galaxies they host. Importantly, while the total $z=0$ HIMF is among the calibration constraints of GAEA2023, its decomposition by host-halo mass is not; the agreement with the \citet{jones20} group measurements therefore provides an independent validation that the model captures not only the distribution of \HI\ masses but also their correlation with the underlying dark-matter halo mass. The agreement is particularly good near the knee of the distributions. At the low-\HI-mass end, residual differences should be interpreted cautiously: on the observational side they reflect ALFALFA completeness and the construction and halo-mass assignment of the group catalog, while the lowest-mass model bins are additionally affected by the finite resolution of the simulations.

\section{The \HI\ in Dark Matter halos: \HI--halo mass relation}
\label{sec:Mh_MHI}

\subsection{The \HI--halo mass relation at \texorpdfstring{$z=0$}{z=0}}
\label{sec:Mh_MHI_z0}

\begin{figure}
	\includegraphics[width=\columnwidth]{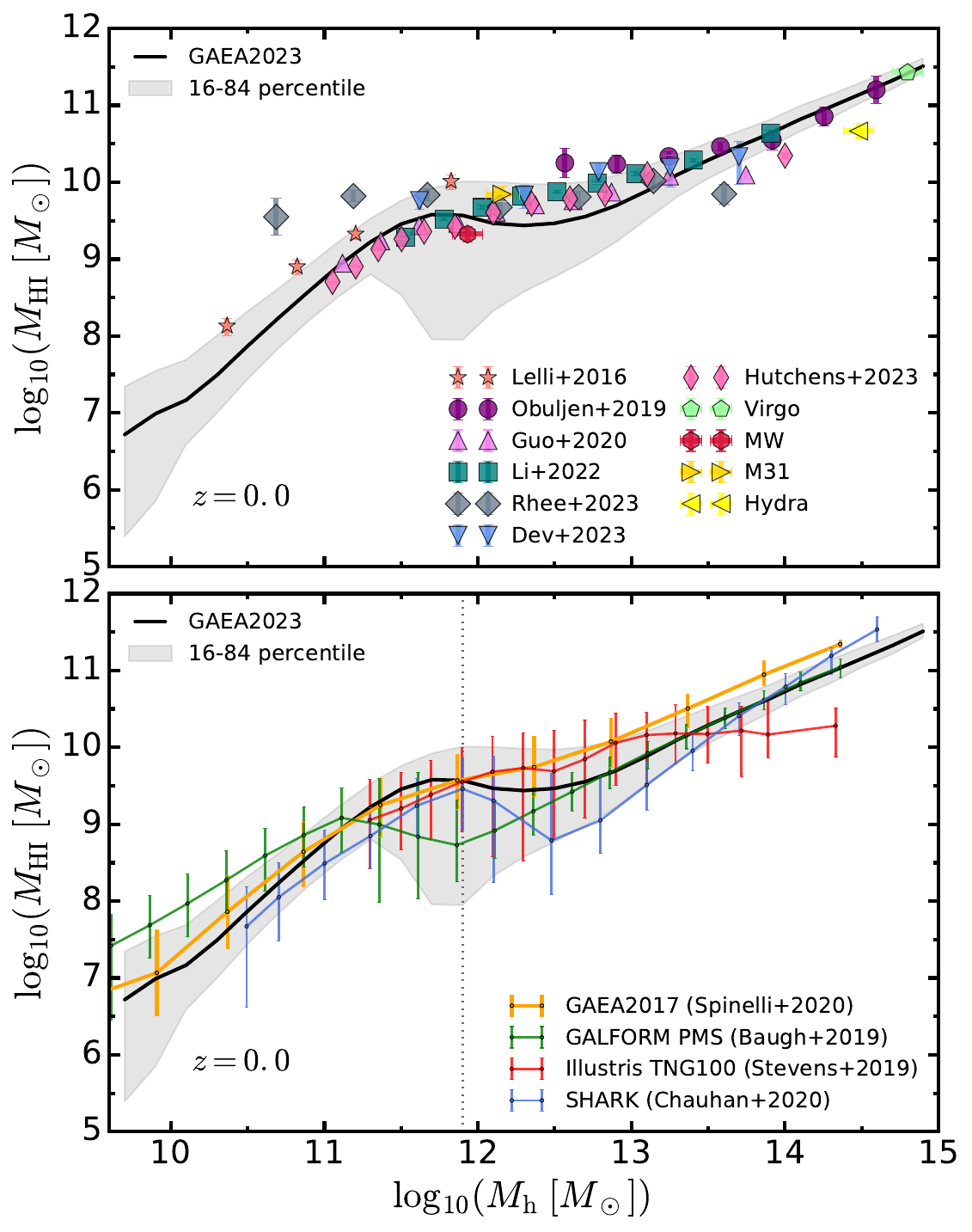}
    \caption{Median \HI--halo mass relation, $M_{\rm HI}(M_{\rm h})$, at $z=0$ for all galaxies within dark matter halos. The black solid curve is the GAEA2023 median; the vertical dotted line (lower panel only) marks the MSII--MSI stitching scale, $\log(M_{\rm h}/M_\odot)\simeq 11.9$. The black shaded band, and the error bars on the model comparison curves, denote 16th--84th percentile ranges at fixed halo mass. \textit{Upper panel:} comparison with observational \HI\ estimates -- salmon stars: individual galaxies \citep{lelli16}; purple circles: ALFALFA group-based \citep{obuljen19}; purple upward triangles: ALFALFA stacking \citep{guo20}; teal squares: \citet{li22}; slate gray diamonds: DINGO/ASKAP stacking \citep{rhee23}; cornflower-blue downward triangles: \citet{dev23}; hot-pink thin diamonds: \citet{hutchens23}; pale-green pentagons: Virgo \citep{li22_virgo}; crimson hexagon: Milky Way \citep{kalberla09}; gold right-pointing triangle: M31 \citep{chemin09}; yellow left-pointing triangle: Hydra \citep{wang21}. \textit{Lower panel:} comparison with other models -- orange: GAEA2017 \citep{spinelli20}; green: GALFORM on the Planck Millennium Simulation \citep{baugh19}; red: IllustrisTNG100 \citep{stevens19}; royal blue: SHARK \citep{chauhan20}.}
    \label{fig:median_hihm_z0}
\end{figure}

Fig.~\ref{fig:median_hihm_z0} presents the $z=0$ \HI--halo mass relation, $M_{\rm HI}(M_{\rm h})$, and compares GAEA2023 to a broad set of theoretical and observational constraints. A halo is identified as a friends-of-friends (FoF) group of virial mass $M_{\rm h}$, and $M_{\rm HI}$ is the total \HI\ mass of its member galaxies -- the central together with all of its satellites and orphans (Sect.~\ref{sec:simulation}) -- so that the relation traces the median halo-summed \HI\ mass as a function of $M_{\rm h}$. Unless otherwise stated, all $M_{\rm HI}(M_{\rm h})$ relations and halo-property maps presented in this work are constructed by stitching the higher-resolution MSII run at low halo mass to the larger-volume MSI run at high halo mass, joined at the halo mass, $\log(M_{\rm h}/M_\odot)\simeq 11.9$, where their median relations converge.

The GAEA2023 prediction traces the median $M_{\rm HI}(M_{\rm h})$. The scatter band indicates substantial halo-to-halo variance at fixed halo mass. At the low-mass end ($\log (M_{\rm h}/M_\odot) \lesssim 11.5$), the relation rises with a shallow slope, reflecting the increasing efficiency of gas accretion and cooling as the halo potential deepens, while stellar feedback continues to regulate the cold gas reservoir. In the intermediate-mass regime ($\log (M_{\rm h}/M_\odot) \sim 11.5$--$12.5$), the relation flattens, reaches a maximum median \HI\ mass near $\log (M_{\rm h}/M_\odot) \sim 11.7$, and then declines toward group scales, consistent with the growing influence of AGN feedback and environmental processes. At the high-mass end ($\log (M_{\rm h}/M_\odot) \gtrsim 12.5$), the median rises again toward cluster scales, likely driven by the cumulative \HI\ of the more massive satellites that retain some gas despite ram-pressure stripping and strangulation (Sect.~\ref{sec:Mh_MHI_diffz}); the numerous low-mass satellites, most strongly affected by these processes, are largely gas-poor and contribute negligibly to the halo \HI\ budget. This picture is consistent with \citet{chen24}, who find that GAEA satellites remain markedly more \HI-rich in cluster environments than their counterparts in hydrodynamical simulations such as IllustrisTNG. This high-mass behavior is set by the larger-volume MSI run, which samples the rare group- and cluster-scale halos. Because the MSI and MSII relations converge near the onset of the intermediate-mass downturn, where the two runs are joined, the handoff is smooth and the high-mass behavior is not an artifact of the MSI--MSII stitching; its amplitude nonetheless remains sensitive to how efficiently environmental stripping is modeled in these dense environments, which sets the \HI\ retained by the surviving satellites. Direct observational constraints at cluster scales are, moreover, sparse, since deep \HI\ observations can be carried out only for the small number of massive clusters in the nearby Universe.

In the upper panel, the GAEA2023 relation passes through the bulk of the observational constraints within their quoted uncertainties. At low and intermediate masses, it is consistent with group-based and stacking estimates from ALFALFA \citep{obuljen19,guo20}, the \HI--halo relation of \citet{li22}, and spectral-stacking results from DINGO/ASKAP \citep{rhee23} and other group-based measurements \citep{dev23,hutchens23}, which recover the average \HI\ content of halos selected by optical or group catalogs. At higher masses, the model also agrees, within the substantial error bars, with measurements for individual systems such as Virgo \citep{li22_virgo}, Hydra \citep{wang21}, the Milky Way \citep{kalberla09}, and M31 \citep{chemin09}, as well as with rotation-curve based estimates from \citet{lelli16}. The width of the GAEA2023 scatter band is comparable to, or somewhat smaller than, the spread among different observational determinations at fixed halo mass, which are themselves affected by systematic uncertainties in halo mass assignment, \HI\ flux calibration, and the use of stacking versus direct detections.

In the lower panel, the earlier GAEA2017 implementation of \citet{spinelli20} tracks the new GAEA2023 relation closely at low and intermediate halo masses, but lies above it for $\log (M_{\rm h}/M_\odot) \gtrsim 12$, with the largest offset ($\Delta \log M_{\rm HI} \sim 0.2$--$0.3$ dex) around the transition region ($\log (M_{\rm h}/M_\odot) \sim 12$--$12.5$) where GAEA2023 dips. In other words, the updated model predicts somewhat lower median \HI\ masses around the transition region, consistent with its revised treatment of satellite stripping, ram-pressure effects, and AGN-driven outflows, although a dedicated comparison would be required to isolate the contribution of each process. The GALFORM predictions from \citet{baugh19} and the SHARK semi-analytic model from \citet{chauhan20} bracket the GAEA2023 relation over most of the mass range: GALFORM yields systematically lower \HI\ content near Milky Way-mass halos ($\log (M_{\rm h}/M_\odot) \sim 12$), while SHARK predicts a higher normalization around the knee, underscoring the sensitivity of $M_{\rm HI}(M_{\rm h})$ to the details of gas partitioning, reincorporation, and feedback prescriptions in different SAMs. The hydrodynamical IllustrisTNG100 prediction \citep{stevens19} broadly follows a similar overall shape to GAEA2023 but exhibits a lower median \HI\ mass at massive halo scales ($\log (M_{\rm h}/M_\odot) \gtrsim 12.5$), possibly reflecting differences in AGN quenching and in the environmental stripping of satellite gas.

\subsection{The \HI--halo mass relation for different \texorpdfstring{$z$}{z}}
\label{sec:Mh_MHI_diffz}
\begin{figure}
	\includegraphics[width=\columnwidth]{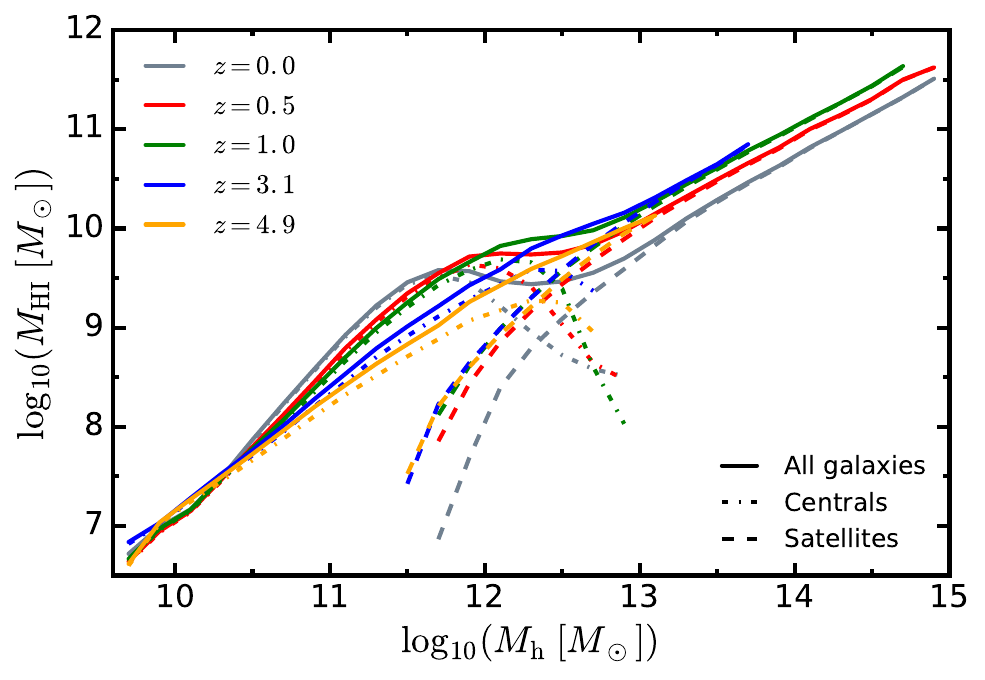}
    \caption{Redshift evolution of the median \HI--halo mass relation in GAEA2023, for all galaxies, centrals, and satellites. The satellite relation is taken from MSI alone, whose larger volume best samples the satellite-dominated, high-mass halos. Solid curves show the median $M_{\rm HI}(M_{\rm h})$ for all galaxies at $z=0$ (slate gray), $z=0.5$ (red), $z=1.0$ (green), $z=3.1$ (blue), and $z=4.9$ (orange). Dash–dotted and dashed curves denote the corresponding relations for central and satellite galaxies, respectively.}
    \label{fig:median_hihm_diff_z}
\end{figure}

Fig.~\ref{fig:median_hihm_diff_z} shows the redshift evolution of the median \HI--halo mass relation in GAEA2023, decomposed into all galaxies, centrals, and satellites. The figure focuses on the regime $\log (M_{\rm h}/M_\odot)\gtrsim 11$, where the central and satellite decompositions are well sampled. With increasing redshift, the \HI\ content in the dip and high-mass regime ($\log (M_{\rm h}/M_\odot)\gtrsim 12$) increases while the low-mass end decreases modestly, so that the pronounced intermediate-mass dip seen at $z=0$ progressively fills in; by $z\gtrsim 3$ the relation approaches an almost single power law, with only a weak inflection between the low- and high-mass ends.

Central galaxies dominate the \HI\ budget at low and intermediate halo masses. At low halo mass, their median \HI\ content increases toward low redshift, while the characteristic turnover shifts to higher halo masses at earlier epochs and its height is non-monotonic in redshift, peaking at $z\simeq 0.5$--$1$. Beyond the turnover, the median \HI\ mass of centrals declines rapidly with increasing halo mass, most steeply at $z\lesssim 1$, consistent with increasingly effective AGN-driven suppression of the cold-gas reservoir in massive centrals. This downturn is much stronger than in GAEA2017 \citep{spinelli20}, consistent with the revised AGN feedback and AGN-driven outflows included in GAEA2023 \citep{fontanot20}.

The halo-summed satellite contribution evolves differently. At fixed halo mass it increases from $z=0$ to $z\simeq 1$ but changes little thereafter, the $z=1$, $z=3.1$, and $z=4.9$ curves lying close together over their common resolved range. Because this quantity is the total \HI\ of all satellites in a halo, rather than that of an individual satellite, its evolution may reflect a combination of more gas-rich satellites at early times, changes in the number and mass distribution of satellites, and the redshift dependence of environmental stripping, which the summed relation alone cannot separate. Relative to GAEA2017 \citep{spinelli20}, GAEA2023 shows a clearer separation between its low- and high-redshift satellite curves, indicating a different balance among satellite assembly, gas supply, and environmental removal \citep{xie20}.

Because the total \HI\ content of massive halos is increasingly dominated by satellites, the satellite evolution is directly imprinted on the all-galaxies relation at the high-mass end: the rise in the satellite contribution between $z=0$ and $z\simeq 1$ drives the corresponding increase in total \HI\ mass, while its near-constancy at $z\gtrsim 1$ accounts for the weak evolution of the total relation. The central and satellite components thus together explain the late emergence of the intermediate-mass dip and its disappearance toward high redshift.

\begin{figure}
    \centering
    \includegraphics[width=\columnwidth]{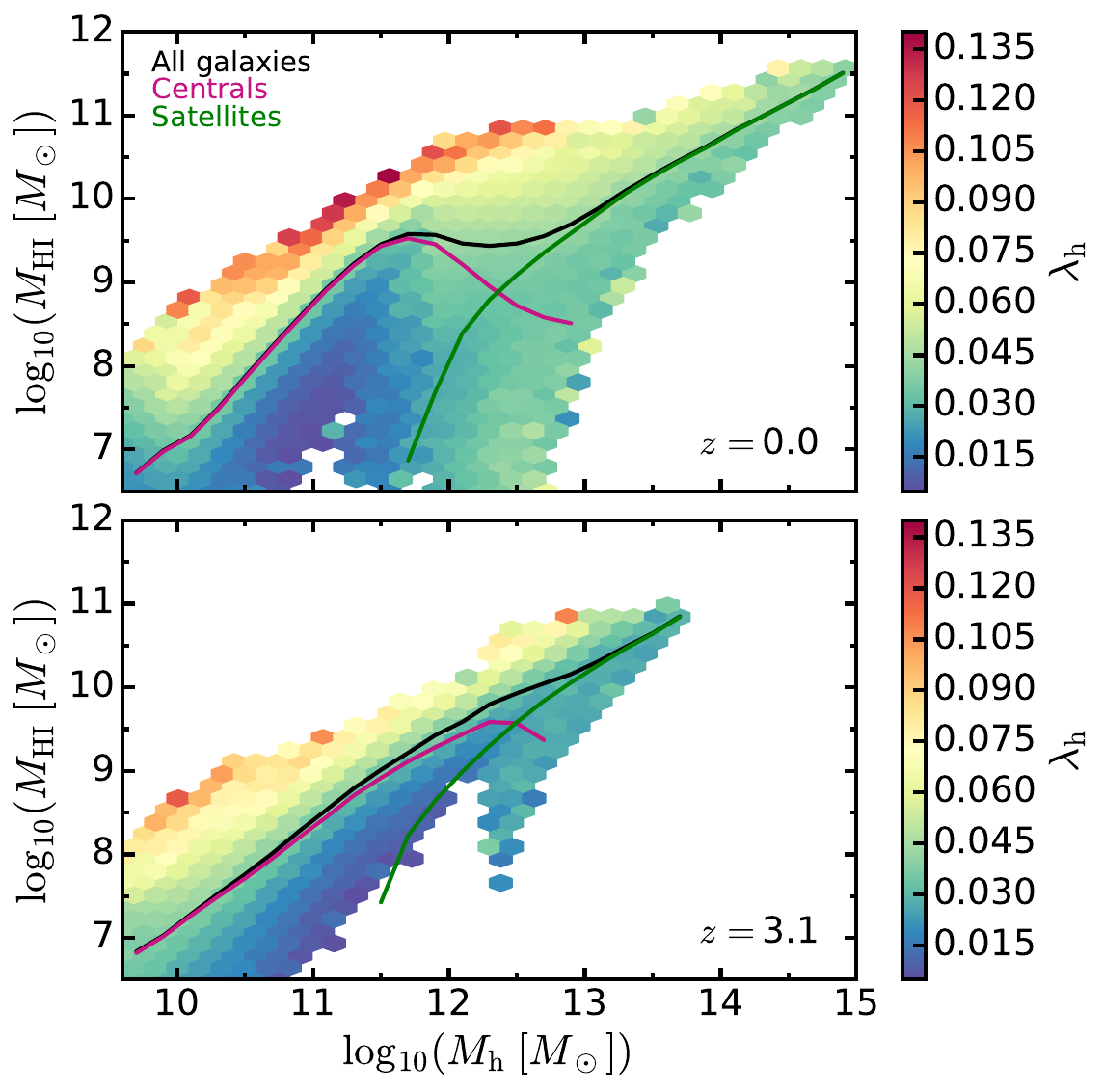}
    \caption{Distribution of the halo spin parameter, $\lambda_{\rm h}$, in the $M_{\rm h}$--$M_{\rm HI}$ plane for GAEA2023 at $z=0$ (upper) and $z=3.1$ (lower). The binned quantity is the total \HI\ mass of each parent halo (centrals plus satellites), and the hexagonal bins are colored by the median $\lambda_{\rm h}$ of the host halos. The overplotted black, magenta, and green solid lines show the median $M_{\rm HI}(M_{\rm h})$ relations for all galaxies, centrals, and satellites, respectively.}
    \label{fig:hi_vs_CentralMvir_vs_spin}
\end{figure}

\begin{figure}
    \centering
    \includegraphics[width=\columnwidth]{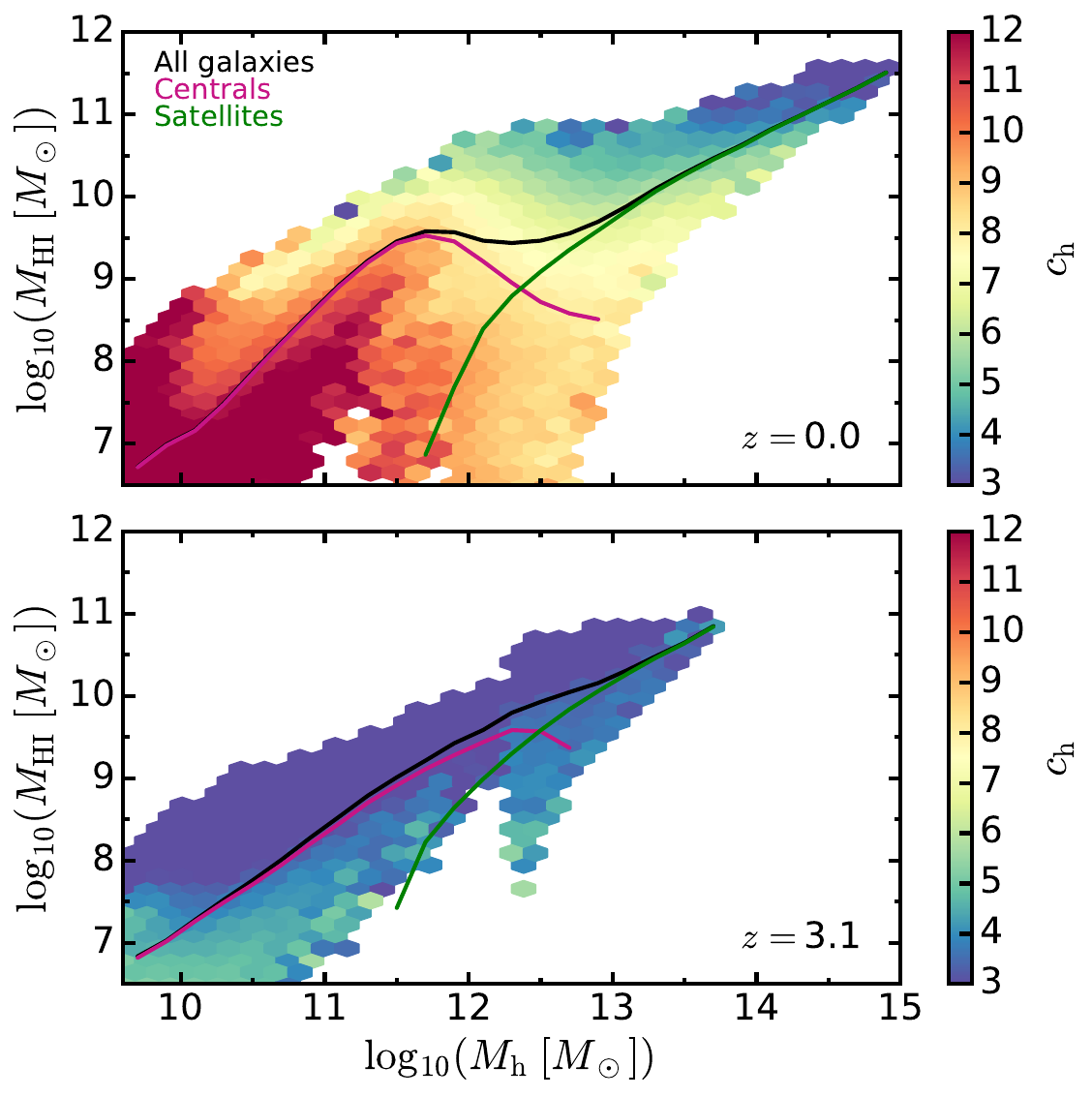}
    \caption{Same as Fig.~\ref{fig:hi_vs_CentralMvir_vs_spin}, but for the median halo concentration, $c_{\rm h}$, of the host halos.}
    \label{fig:hi_vs_CentralMvir_vs_conc}
\end{figure}

\section{Interpreting the \texorpdfstring{\HI}{HI}--halo mass relation in GAEA2023}
\label{sec:physical drivers}

The shape of the $M_{\rm HI}(M_{\rm h})$ relation provides only a partial description of the connection between \HI\ and dark-matter halos. To characterize this scaling in a way that is useful for modeling and for mock-catalog construction, it is equally important to quantify the associated scatter and to understand its physical origin. We therefore investigate how the dispersion in $M_{\rm HI}$ at fixed $M_{\rm h}$ correlates with secondary halo properties that are directly accessible in dark-matter-only simulations, such as the halo spin parameter, the concentration, and the assembly history.

A complete physical interpretation of the scatter in $M_{\rm HI}$ at fixed $M_{\rm h}$ is challenging: it would require decomposing the halo \HI\ budget into central and satellite contributions across redshift, and disentangling how cooling suppression, AGN feedback, environmental stripping, and the atomic/molecular partition jointly regulate it over the full halo-mass range. In the main text we therefore focus on intrinsic halo properties -- principally the spin parameter $\lambda_{\rm h}$ and the concentration $c_{\rm h}$ -- that correlate most directly with the \HI\ content of the central galaxy, and hence with the total \HI\ budget of the halo, and that, being directly available in dark-matter-only catalogs, are particularly useful for populating such simulations with \HI. Additional galaxy- and baryon-level properties are examined in Appendix~\ref{sec:physical drivers cont}; these provide complementary physical insight but do not constitute an exhaustive description of all the processes governing the scatter. The halo properties considered below are not independent, and are examined separately only to characterize their individual correlations with the \HI--halo mass relation.

\subsection{The halo spin parameter (\texorpdfstring{$\lambda_{\rm h}$)}{lambda}}
\label{sec:spin parameter}
The (dimensionless) halo spin parameter provides a convenient measure of the specific angular-momentum content of dark-matter halos and is expected to influence the size and surface-density profile of the gaseous discs that form within them. We adopt the Peebles-type definition \citep{peebles69},
\begin{equation}
\lambda_{\rm h} = \frac{J_{\rm h}\,|E_{\rm h}|^{1/2}}{G\,M_{\rm h}^{5/2}} \,,
\end{equation}
where $J_{\rm h}$ is the total angular momentum of the halo, $E_{\rm h}$ its total (kinetic plus potential) energy, $M_{\rm h}$ the halo virial mass, and $G$ the gravitational constant. For each halo, $\lambda_{\rm h}$ is computed by the halo finder from the dark-matter particle distribution and stored in the merger-tree catalog, from which we take it directly. This parametrization has been widely used in $N$-body and hydrodynamical simulations, which show that dark-matter halos follow an approximately lognormal spin distribution with only a weak dependence on mass and environment \citep{Bullock2001, zjupa16}. Hydrodynamical simulations further indicate that the baryonic component can attain a higher specific angular momentum than the dark matter \citep{stewart13, zjupa16}, an enhancement that is most pronounced for gas accreted through cold streams \citep{danovich15}. The magnitude of this offset remains uncertain, however, and may depend on the gas-accretion mode, feedback prescriptions, and resolution; moreover, halo and baryonic angular momenta are correlated but not identical, with substantial scatter introduced by accretion, feedback, and mergers.

In analytic disc-formation models, higher-spin halos are predicted to host more extended, lower-surface-density gas discs that convert gas into stars less efficiently and therefore tend to retain larger cold-gas and \HI\ reservoirs; conversely, low-spin halos form more compact, higher-surface-density discs that are more susceptible to rapid star formation and feedback-driven gas removal \citep{dalcanton97, mo98}. These expectations are borne out qualitatively in several galaxy-formation studies. Semi-analytic work with the SHARK model finds that \HI-rich systems preferentially reside in higher-spin halos and that halo spin acts as an important secondary parameter controlling the radial extent of \HI\ discs \citep{chauhan20}. Within the same GAEA framework adopted here, \citet{zoldan18, zoldan19} show that the explicit evolution of the angular momentum of the gaseous and stellar discs reproduces the observed galaxy size--mass and specific-angular-momentum relations. These results indicate that GAEA captures the link between angular momentum and disc structure realistically, providing physical support for the spin dependence of \HI\ content examined here.

On the observational side, direct measurements of halo spin for individual galaxies remain challenging. Using a semi-analytic analysis of \HI-bearing ultra-diffuse galaxies, \citet{rong24a} found that these systems have higher stellar and gaseous specific angular momenta than typical dwarf galaxies, and argued that their high gas specific angular momentum helps explain their elevated gas fractions and low star-formation efficiencies. This interpretation is qualitatively consistent with the angular-momentum-based disc evolution implemented in GAEA, although it does not constitute a direct observational measurement of the halo-spin--\HI\ relation. Building on this work, \citet{liu25} estimated halo spins statistically from the observed stellar masses, sizes, and \HI\ contents of ALFALFA galaxies and found a positive correlation between halo spin and the \HI-to-stellar mass ratio across a broad mass range. While these empirical inferences remain model-dependent, their trends are consistent with analytic disc-formation arguments and with the positive association between halo spin and \HI\ content found in the SHARK \citep{chauhan20} and GAEA models.

Motivated by these considerations, we treat $\lambda_{\rm h}$ as a key secondary parameter of the $M_{\rm HI}(M_{\rm h})$ relation. Fig.~\ref{fig:hi_vs_CentralMvir_vs_spin} shows its distribution in the $M_{\rm h}$--$M_{\rm HI}$ plane. At fixed halo mass, halos with higher total \HI\ content tend to have systematically larger median spin parameters, particularly below the halo mass at which the median \HI\ content peaks. This suggests that, in GAEA2023, variations in halo angular momentum contribute to the scatter in halo \HI\ content as long as the total \HI\ budget remains closely tied to the central galaxy (see Appendix \ref{sec:spin parameter cont} for more details), and is qualitatively consistent with the SHARK results of \citet{chauhan20}. At the highest halo masses, where the median relation flattens and then rises owing to the growing satellite contribution, the variation of $\lambda_{\rm h}$ across the \HI\ range becomes much weaker; spin therefore appears to play a progressively smaller role in setting the total \HI\ content once halos reach group and cluster scales. Since halo spin is itself correlated with other structural properties such as concentration and assembly history, the trends presented here should be interpreted as correlations rather than as evidence for a unique causal role of spin.

We caution that GAEA explicitly follows the angular-momentum evolution of the gaseous and stellar discs \citep{xie20}, while the atomic/molecular partition depends on the resulting disc surface density \citep{xie17}. A positive association between halo spin and \HI\ content is therefore partly encoded in the model framework and should not be regarded as a fully independent prediction. SHARK likewise follows the exchange of angular momentum among its baryonic components and employs a surface-density-dependent partition of atomic and molecular gas \citep{lagos18,chauhan20}. The agreement between the two models therefore shows that the trend is recovered by distinct semi-analytic implementations, but does not provide an entirely independent test of the underlying physical mechanism; its strength, halo-mass dependence, and redshift evolution remain non-trivial model predictions.

\subsection{The halo concentration (\texorpdfstring{$c_{\rm h}$)}{ch}}
\label{sec:halo concentration}

A second halo property that may regulate the scatter of the $M_{\rm HI}(M_{\rm h})$ relation is the concentration of the dark-matter density profile. Cosmological $N$-body simulations show that the spherically averaged profiles of relaxed halos are well described by the two-parameter form of \citet{NFW1996, NFW1997} (hereafter NFW),
\begin{equation}
\rho(r) = \frac{\rho_{\rm s}}{(r/r_{\rm s})\,(1 + r/r_{\rm s})^{2}},
\end{equation}
where $r_{\rm s}$ is the scale radius and $\rho_{\rm s}$ the characteristic density. At fixed halo mass the profile is then fully specified by the single shape parameter
\begin{equation}
c_{\rm h} \equiv \frac{R_{\rm vir}}{r_{\rm s}},
\end{equation}
the concentration, defined as the ratio of the virial radius to the scale radius. Larger $c_{\rm h}$ corresponds to a more centrally concentrated mass distribution and a deeper inner potential well.

Rather than fitting an NFW density profile to each halo individually\footnote{Directly fitting an NFW profile requires access to the spatial distribution of the dark-matter particles within each halo. While particle data are available for the $z=0$ snapshot, they are not available for all simulation outputs spanning the redshift range analyzed in this work.}, we recover $c_{\rm h}$ from the two characteristic circular velocities provided directly by the halo finder: the virial circular velocity $V_{\rm vir} = (G\,M_{\rm h}/R_{\rm vir})^{1/2}$ and the peak of the circular-velocity curve, $V_{\rm max}$. For an NFW halo these obey the relation \citep[e.g.][]{Springel2008, Klypin2011, Prada2012}
\begin{equation}
\left(\frac{V_{\rm max}}{V_{\rm vir}}\right)^{2} =
0.2162\,\frac{c_{\rm h}}{\ln(1+c_{\rm h}) - c_{\rm h}/(1+c_{\rm h})},
\label{eq:vmaxvvir}
\end{equation}
which follows from the NFW circular-velocity curve attaining its maximum at $r_{\rm max} \simeq 2.16\,r_{\rm s}$. We invert Eq.~\ref{eq:vmaxvvir} numerically for every central halo to obtain its concentration.\footnote{$V_{\rm max}/V_{\rm vir}$ is a non-monotonic function of $c_{\rm h}$ with a minimum at $c_{\rm h}\simeq 2.16$; we restrict the inversion to the physical, monotonically increasing branch $c_{\rm h}\gtrsim 2.16$.} Because $V_{\rm max}$ is measured directly from the circular-velocity curve, while $V_{\rm vir}$ derives from the halo virial mass $M_{\rm h}\,(\equiv M_{\rm vir})$ and radius $R_{\rm vir}$, both are robust halo-finder outputs; this approach yields $c_{\rm h}$ without per-halo profile fits and is well suited to the large halo samples considered here. It nonetheless assumes that the halos are dynamically relaxed and well described by an NFW profile, so the inferred $c_{\rm h}$ should be regarded as approximate for recently merged or unrelaxed systems. The MSII--MSI stitching introduced in Sect.~\ref{sec:Mh_MHI_diffz} is particularly important for this property, since MSI under-resolves the inner profiles -- and hence $c_{\rm h}$ -- of halos below $\log_{10}(M_{\rm h}/M_\odot)\sim 11.5$.

Concentration is of particular interest because it retains information about halo assembly. Halos that assemble the bulk of their mass earlier, when the mean density of the Universe was higher, develop higher characteristic densities and therefore larger concentrations \citep{NFW1997,Bullock2001,Wechsler2002,Ludlow2014}. At fixed halo mass, $c_{\rm h}$ can thus be used as a proxy for formation epoch and is closely related to the formation redshift $z_{50}$, defined as the redshift by which a halo had assembled half of the mass it has at the epoch of interest. Concentration is consequently one of the canonical secondary halo properties associated with assembly bias \citep{Gao2005,Wechsler2006}. Later-forming, less-concentrated halos may retain larger atomic-gas reservoirs at fixed mass because they have had less time to consume or expel their cold gas and may continue accreting it over a longer interval. This interpretation, together with a direct analysis based on $z_{50}$, is discussed in more detail in Appendix~\ref{sec:halo formation history}. Examining the scatter in $M_{\rm HI}(M_{\rm h})$ as a function of $c_{\rm h}$ therefore provides a catalog-level test of the assembly dependence inferred from the formation-time analysis.

Fig.~\ref{fig:hi_vs_CentralMvir_vs_conc} shows the $c_{\rm h}$ distribution in the same plane. At $z=0$ (upper panel) it displays two distinct gradients. First, the median concentration declines steeply with halo mass, from $c_{\rm h}\sim 10$--$12$ at $\log_{10}(M_{\rm h}/M_\odot)\sim 10$ to $c_{\rm h}\sim 3$--$5$ at $\log_{10}(M_{\rm h}/M_\odot)\gtrsim 13$, recovering the well-known decrease of concentration with mass \citep[e.g.][]{DuttonMaccio2014, DiemerJoyce2019}. Second, and more relevant here, a clear vertical gradient is present at fixed halo mass: \HI-richer halos are systematically hosted by less concentrated systems than \HI-poorer halos of the same mass. This stratification is most pronounced around and above the characteristic mass $\log_{10}(M_{\rm h}/M_\odot)\sim 12$, where the median central relation turns over and begins to decline, and it persists into the satellite-dominated regime at higher masses, where the most \HI-rich halos define the least-concentrated upper envelope. The sense of the trend -- lower $c_{\rm h}$ at higher \HI\ content -- is consistent with later-forming halos retaining more atomic gas, and it mirrors the dependence on $z_{50}$ found in Appendix~\ref{sec:halo formation history}, supporting the interpretation that concentration captures the same assembly-driven component of the scatter.

By $z=3.1$ (lower panel), the concentration range has narrowed substantially, with most of the plane spanning $c_{\rm h}\sim3$--$5$, consistent with the decrease of halo concentration toward earlier epochs at fixed mass \citep{DuttonMaccio2014,DiemerJoyce2019}. Given this limited dynamic range, the vertical concentration gradient at fixed halo mass is much weaker than at $z=0$: only a mild tendency remains for the \HI-poorer lower envelope to be more concentrated than the \HI-rich ridge. Concentration is therefore a less effective discriminator of the \HI\ scatter at this epoch. The more pronounced variation visible in the spin map suggests that halo spin may provide more information about the high-redshift scatter, although this comparison is qualitative and the relative predictive power of the two properties is quantified only through the fitting analysis in Sect.~\ref{sec:hihm_scatter_fit}. Overall, concentration is most strongly correlated with $M_{\rm HI}$ at low redshift, particularly around and above the characteristic halo-mass scale, while this correlation weakens toward higher redshift.

As for the halo spin, these trends should be read as correlations rather than as evidence for a unique causal role of concentration, since concentration, spin, and formation time are themselves mutually correlated. Unlike halo spin, concentration is not an explicit input to the GAEA disc model. Its association with \HI\ content is therefore an emergent correlation within the model, most plausibly reflecting the shared dependence of halo structure and of gas-accretion and consumption histories on halo assembly, rather than a dependence imposed directly by the disc prescriptions.

\section{Fitting the \HI\ halo mass relation}
\label{sec:hihm_fit}

\subsection{Fitting the median \HI\ halo mass relation}
\label{sec:hihm_median_fit}

\begin{figure}
\centering
\includegraphics[width=\columnwidth]{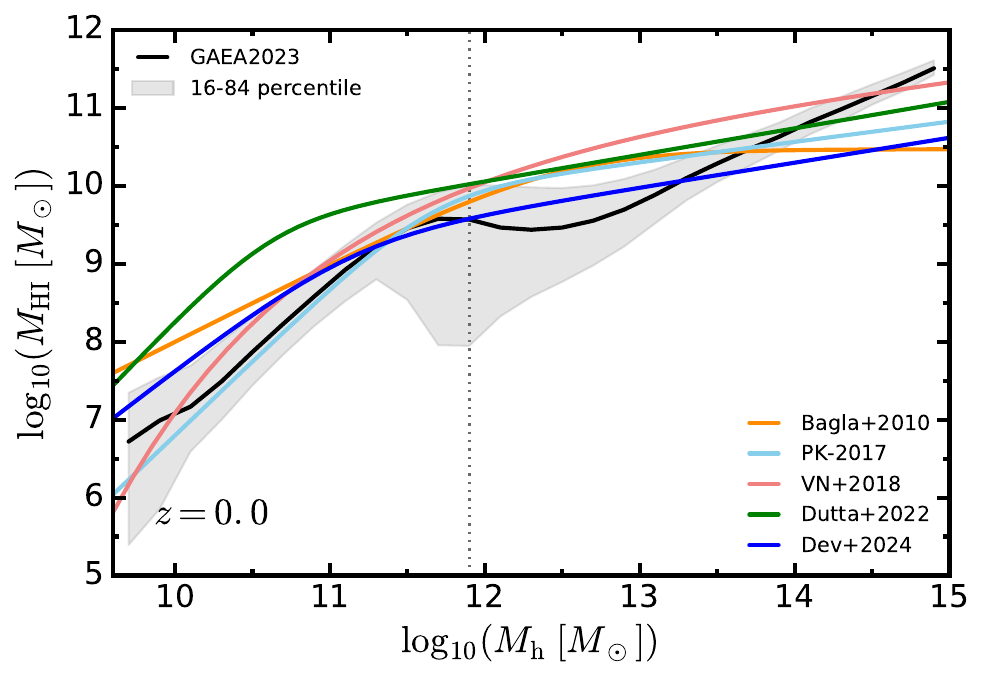}
\caption{Same as Fig.~\ref{fig:median_hihm_z0}, but here GAEA2023 median is compared with analytic and empirical $M_{\rm HI}(M_{\rm h})$ prescriptions -- dark orange: \citet{bagla10}; sky blue: \citet{padmanabhan17}; light coral: \citet{Villaescusa-Navarro:2018vsg}; green: \citet{dutta22}; blue: \citet{dev24}.}
\label{fig:hihm_pre_model}
\end{figure}

\begin{figure}
\centering
\includegraphics[width=\columnwidth]{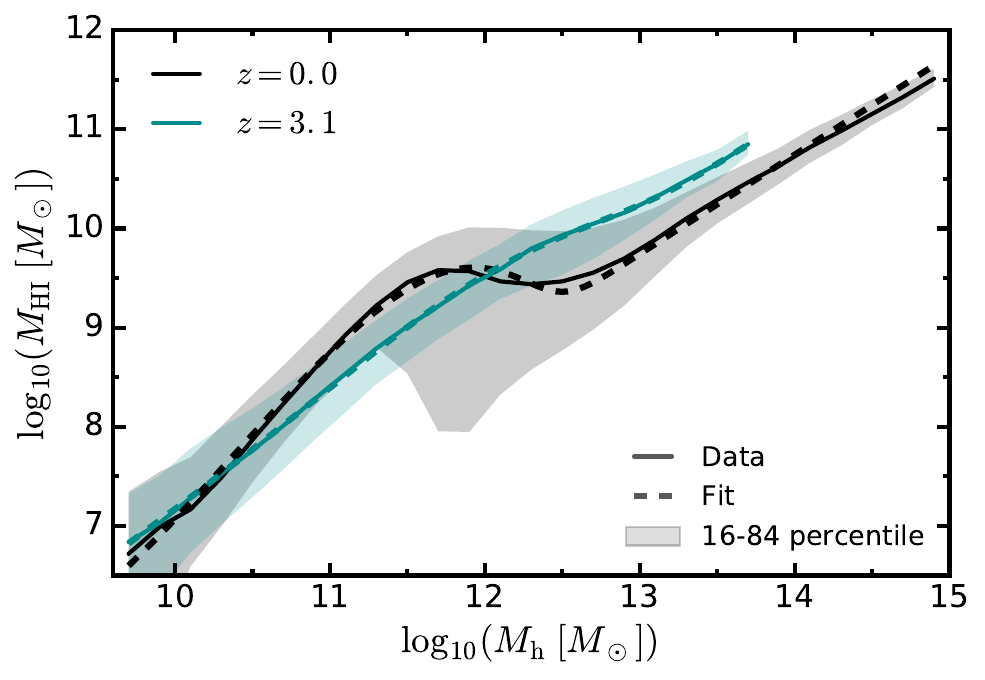}
\caption{Total \HI--halo mass relation, $M_{\rm HI}(M_{\rm h})$ (central plus satellite \HI\ per parent halo), for GAEA2023 at $z=0$ (black) and $z=3.1$ (dark cyan). For each redshift, the solid line shows the GAEA median and the dashed line the best-fitting Eq.~\ref{eq:hihm} (Table~\ref{tab:hihm}). Shaded bands indicate the 16th--84th percentile scatter.}
\label{fig:hihm_two_z}
\end{figure}

\begin{table}
\centering
\setlength{\tabcolsep}{4pt}
\caption{Best-fitting parameters of the total (central\,$+$\,satellite) \HI--halo mass relation $M_{\rm HI}(M_{\rm h})$, Eq.~\ref{eq:hihm}, for the GAEA2023 snapshots closest to $z=0$--$5$. Masses are in physical $M_\odot$ ($h=0.73$); $a_1$ and $a_2$ are dimensionless. The low-mass cut-off $M_{\rm min}=10^{8}\, M_\odot$ and the exponent $\gamma=0.5$ are held fixed. $^{\dagger}$\,$M_{\rm break}$ (and, by $z=4.9$, $a_2$) are unconstrained: the fit rails $M_{\rm break}$ at the lower bound of its prior and these values are not to be interpreted physically (see text).}
\label{tab:hihm}
\begin{tabular}{lccccc}
\hline\hline
$z$ & $a_1$ & $a_2$ & $\alpha$ & $\beta$ & $\log_{10}(M_{\rm break}/M_\odot)$ \\
\hline
0.0 & $1.3\times10^{-3}$ & $5.5\times10^{-4}$ & 0.63 & 1.17 & $11.03$ \\
1.0 & $7.0\times10^{-4}$ & $1.1\times10^{-3}$ & 0.44 & 1.32 & $10.68$ \\
2.1 & $5.0\times10^{-4}$ & $1.5\times10^{-3}$ & 0.40 & 1.22 & $10.66$ \\
3.1 & $6.5\times10^{-4}$ & $1.4\times10^{-3}$ & 0.35 & 1.05 & $10.50^{\dagger}$ \\
3.9 & $1.2\times10^{-3}$ & $1.1\times10^{-3}$ & 0.31 & 0.75 & $10.50^{\dagger}$ \\
4.9 & $3.4\times10^{-3}$ & $\sim 0^{\dagger}$ & 0.23 & 0.41 & $10.50^{\dagger}$ \\
\hline
\end{tabular}
\end{table}

The \HI--halo mass relation, $M_{\rm HI}(M_{\rm h})$, is a central ingredient of halo-model and halo-occupation descriptions of the post-reionization 21-cm signal (Sect.~\ref{sec:intro}). It has been characterized using empirical and halo-model parametrizations \citep{bagla10,BarnesHaehnelt2010,santos2015,padmanabhan_refregier17,obuljen19}, hydrodynamical simulations \citep{crain17,Villaescusa-Navarro:2018vsg,diemer19,stevens19}, and semi-analytic models \citep{baugh19,spinelli20,chauhan20}. These approaches generally predict that \HI\ is suppressed in low-mass halos by photoionization and stellar feedback, but differ substantially in the transition and high-mass regimes, where AGN feedback and the satellite population become important. Figure~\ref{fig:hihm_pre_model} compares representative published prescriptions with the GAEA2023 median relation; the predictions span a broad range at the low- and high-mass ends, whereas the more recent relations of \citet{dutta22} and \citet{dev24} track the GAEA2023 result relatively closely over the halo masses constrained by current data.

Our purpose is not simply to add another median relation to the literature, but to provide a self-consistent, redshift-dependent prescription for both the median and its dependence on secondary halo properties. Although the local HIMF is among the calibration constraints of GAEA2023, the detailed dependence of \HI\ content on halo mass and the associated scatter were not directly calibrated. We therefore fit the median relation in this subsection and model its scatter in Sect.~\ref{sec:hihm_scatter_fit}. The resulting prescription summarizes the GAEA2023 predictions in a compact form and allows dark-matter-only catalogs to be populated with \HI\ at low computational cost for the construction of 21-cm mock catalogs. This type of application has already been demonstrated by \citet{spinelli22}, who populated PINOCCHIO halo light cones with \HI\ using an \HI--halo mass relation and its scatter to build end-to-end mock intensity-mapping observations for an SKAO/MeerKAT foreground-cleaning challenge. Separate fits to the central and satellite contributions, for applications that treat these populations independently, are provided in Appendix~\ref{sec:censat_fit}.

We adopt the functional form introduced by \citet{spinelli20}, who extended the \citet{baugh19} parametrization with an explicit low-mass cut-off: 
\begin{equation}
M_{\rm HI}(M_{\rm h}) = M_{\rm h} \left[\, a_1 \left(\frac{M_{\rm h}}{10^{10}\,h^{-1}M_\odot}\right)^{\beta} e^{-\left(M_{\rm h}/M_{\rm break}\right)^{\alpha}} + a_2 \right] e^{-\left(M_{\rm min}/M_{\rm h}\right)^{\gamma}}.
\label{eq:hihm}
\end{equation}
The term in square brackets is the sum of a rising `cooling' branch -- which behaves as $M_{\rm HI}\propto M_{\rm h}^{1+\beta}$ at intermediate masses and is exponentially truncated above $M_{\rm break}$ with sharpness $\alpha$ -- and a constant $a_2$ that drives $M_{\rm HI}\propto M_{\rm h}$ at the high-mass end, the regime where the total \HI\ budget is dominated by satellites. The final factor imposes the low-mass cut-off at $M_{\rm min}$. Following \citet{spinelli20} we fix $\gamma=0.5$. The cut-off scale $M_{\rm min}$ lies below the resolved halo-mass range of our samples and is not constrained by the fit \citep[in][it railed to unphysically small values for $z\ge1$]{spinelli20}; we therefore hold it fixed at $M_{\rm min}=10^{8}\,M_\odot$, leaving five free parameters ($a_1, a_2, \alpha, \beta, M_{\rm break}$). Equation~\ref{eq:hihm} is fit to the stitched MSII$+$MSI median relation, with parameter uncertainties estimated by bootstrap resampling of the binned medians.

We have verified that the fit is insensitive to the halo mass at which the MSII and MSI medians are stitched. This stitching scale lies well above $M_{\rm break}$ at every redshift -- by $\sim0.9$--$1.6$ dex where $M_{\rm break}$ is constrained -- so the break is set entirely by the better-resolved MSII data below the join. Displacing the stitching mass by one bin ($\pm0.2$ dex) about its adopted value changes the recovered $M_{\rm break}$ by less than $0.1$ dex at $z=0$ and $z=1$, well within its bootstrap uncertainty, and by $\sim0.3$ dex at $z=2.1$, still within the (larger) error at that redshift. At $z\gtrsim3$, where $M_{\rm break}$ is already unconstrained (Table~\ref{tab:hihm}), the stitching scale is immaterial. 

The best-fitting parameters are listed in Table~\ref{tab:hihm} for the GAEA2023 snapshots closest to $z=0$--$5$. Figure~\ref{fig:hihm_two_z} compares the GAEA median relation (solid lines, with the 16--84th percentile scatter shaded) with the fit (dashed lines) at $z=0$ and $z=3.1$; the single functional form of Eq.~\ref{eq:hihm} reproduces the median to better than $\sim0.1$--$0.2$ dex over the entire resolved range at every redshift. At $z=0$ (black) the relation displays the full structure outlined above: $M_{\rm HI}$ rises steeply with halo mass, reaches a broad maximum near $M_{\rm h}\sim10^{11.7}\,M_\odot$, declines weakly as the central galaxies are quenched, and turns up again toward group and cluster scales as satellite \HI\ comes to dominate. In Eq.~\ref{eq:hihm} this corresponds to a steep rising branch ($\beta\simeq1.2$, i.e.\ $M_{\rm HI}\propto M_{\rm h}^{2.2}$ before the truncation) cut off at $M_{\rm break}\simeq10^{11}\,M_\odot$, together with the high-mass linear term, whose amplitude $a_2\simeq5\times10^{-4}$ sets the asymptotic ratio $M_{\rm HI}/M_{\rm h}$ in massive halos. We stress that $M_{\rm break}$ is the $e$-folding scale of the cooling branch and \emph{not} the location of the visible maximum: because the satellite term partially fills in the post-peak decline, the apparent turnover lies $\sim0.7$ dex above $M_{\rm break}$. With this caveat, the recovered $M_{\rm break}\sim10^{11}\,M_\odot$ marks the onset of the high-mass suppression, consistent with the halo-mass scale at which AGN feedback becomes efficient in GAEA2023.

By $z=3.1$ (dark cyan) the relation is, to within the scatter, a single near-linear power law, $M_{\rm HI}\propto M_{\rm h}$, over the whole resolved range: neither the high-mass turnover and dip nor the satellite upturn is present, because the massive quenched centrals and rich satellite systems that produce them have not yet assembled. As a consequence the parameters that describe those features -- $M_{\rm break}$, the high-mass amplitude $a_2$, and the rising-branch slope $\beta$, which trade off against the truncation -- are not independently constrained at $z\gtrsim3$; the fit rails $M_{\rm break}$ against the lower edge of its prior (Table~\ref{tab:hihm}). We report these values for completeness but caution that they should not be interpreted physically at high redshift. The overall trend with redshift is thus one of an increasingly \emph{featured} relation toward low $z$: a well-defined break and a satellite upturn emerge only once AGN quenching and satellite assembly have had time to operate, while at early epochs a simpler power-law description is adequate \citep[in line with][]{baugh19, spinelli20}.


\subsection{The scatter of the \HI--halo mass relation: spin and concentration}
\label{sec:hihm_scatter_fit}

Beyond its median, the total \HI--halo mass relation retains a substantial scatter that itself carries information about halo assembly. We model this scatter as a function of the halo spin parameter $\lambda$ \citep{peebles69} and the concentration $c$ (henceforth, $\lambda \equiv \lambda_{\rm h}$ and $c \equiv c_{\rm h}$ for clarity), the latter recovered by inverting the $V_{\rm max}/V_{\rm vir}$--$c$ relation for an NFW profile (Sect.~\ref{sec:halo concentration}; \citealt{Prada2012}). For every host we measure how far its \HI\ content lies above or below the median relation (Eq.~\ref{eq:hihm}) at its halo mass. We call this the deviation from the median,
\begin{equation}
\Delta \equiv \log_{10} M_{\rm HI}
      - \left\langle \log_{10} M_{\rm HI}\right\rangle\!(M_{\rm h}) ,
\label{eq:delta_c}
\end{equation}
so that $\Delta>0$ marks a halo that is \HI-richer than typical for its mass and $\Delta<0$ one that is \HI-poorer, and, at each redshift, we ask how this deviation depends on spin and concentration by fitting a single plane to all halos,
\begin{equation}
\Delta = C + A_{\lambda}\,x_{\lambda} + B_{c}\,x_{c},
\label{eq:plane_c}
\end{equation}
where each predictor is standardized,
\begin{equation}
x_{\lambda}=\frac{\lambda-\tilde{\lambda}}{s_{\lambda}}, \qquad
x_{c}=\frac{c-\tilde{c}}{s_{c}},
\label{eq:standardise_c}
\end{equation}
that is, centered on its typical value $\tilde{\cdot}$ (the median) and divided by its spread $s$ (half the width of the 16--84 percentile range) at that redshift. This places spin and concentration on the same dimensionless footing, so that $A_{\lambda}$ and $B_{c}$ measure how much the deviation $\Delta$ (in dex) changes when the corresponding property increases by one typical spread, and can be compared directly. The intercept $C$ ($|C|\lesssim0.15$~dex at all redshifts) is a normalization -- the mean offset between the sample and the median fit -- and carries no dependence on halo properties. In application the deviation is referenced to the median relation, i.e.\ $C=0$. Because $\lambda$ and $c$ are taken directly from the halo catalogs and carry no observational uncertainties, we treat them as fixed (error-free) predictors and solve Eq.~(\ref{eq:plane_c}) by volume-weighted least squares in $\Delta$. We define $\sigma_{\rm int}$ as the robust dispersion of the residuals about the fitted plane -- half the $16$--$84$ percentile width in $\log_{10} M_{\rm HI}$, which is insensitive to the bimodal tail discussed below. It quantifies the part of the \HI\ scatter not captured by the linear dependence on spin and concentration, and can include unmodeled halo or baryonic properties, nonlinear dependences, and numerical limitations. This differs from the \textsc{hyper-fit} hyperplane \citep{Robotham2015} adopted by \citet{chauhan20}, which minimizes the perpendicular scatter and allows for uncertainties in all variables; expressed as a dispersion in $\log_{10} M_{\rm HI}$, the two agree to within $\lesssim2\%$. Table~\ref{tab:scatter_c} lists the coefficients $A_{\lambda}, B_{c}$, the standardization constants $\tilde\lambda, s_{\lambda}, \tilde c, s_{c}$, and the residual scatter $\sigma_{\rm int}$; the coefficients are effective, number-weighted averages over halo mass\footnote{Because MSII samples a $125\times$ smaller volume than MSI, the pooled fits weight each MSII halo by $V_{\rm MSI}/V_{\rm MSII}=125$, so that the two simulations contribute with a single, volume-consistent number weighting.}, whose mass dependence we discuss below.

\begin{figure}
\centering
\includegraphics[width=\columnwidth]{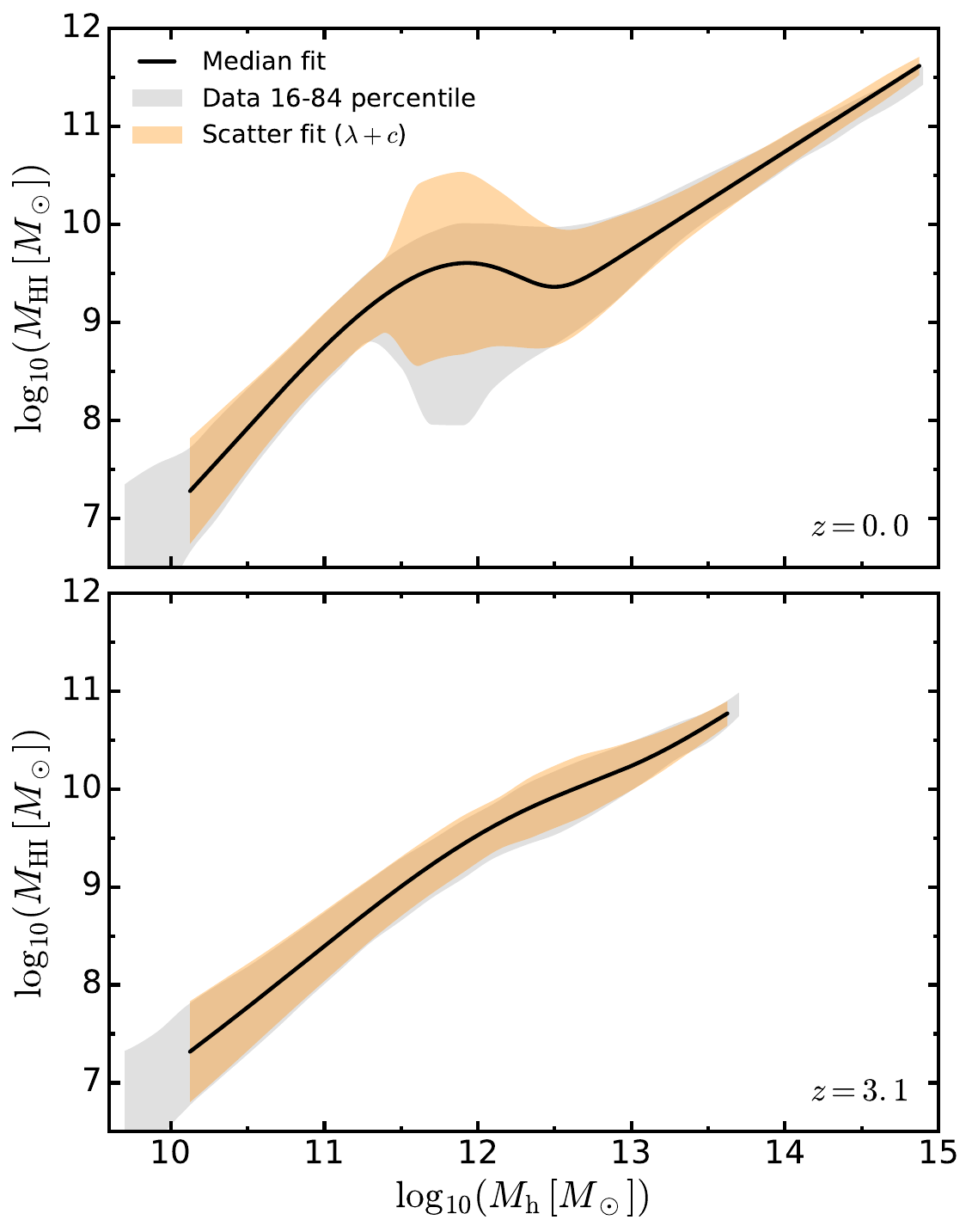}\caption{Total \HI--halo mass relation and its scatter at $z=0$ (upper) and $z=3.1$ (lower). The solid line is the median fit. The gray band is the 16--84 percentile scatter measured in the simulation; the colored band is the scatter produced by our fit (Eq.~\ref{eq:apply_c}) -- the median plus the spin and concentration terms and the intrinsic scatter $\sigma_{\rm int}$.}
\label{fig:scatter_c}
\end{figure}

Equations~(\ref{eq:delta_c})--(\ref{eq:standardise_c}) and Table~\ref{tab:scatter_c} give a complete, self-contained prescription for adding \HI\ scatter to a dark-matter halo catalog. For a halo of mass $M_{\rm h}$, spin $\lambda$, and concentration $c$ at redshift $z$,
\begin{equation}
\log_{10} M_{\rm HI} = \left\langle \log_{10} M_{\rm HI}\right\rangle\!(M_{\rm h}) + A_{\lambda}\,x_{\lambda} + B_{c}\,x_{c} + \mathcal{N}\!\left(0,\sigma_{\rm int}^{2}\right),
\label{eq:apply_c}
\end{equation}
where the median $\left\langle \log_{10} M_{\rm HI}\right\rangle\!(M_{\rm h})$ is given by Eq.~(\ref{eq:hihm}) and Table~\ref{tab:hihm}. The scatter about the median is physical -- it reflects the varied assembly and baryonic histories of individual halos -- rather than random noise; but with only a halo catalog, only the part that correlates with the available properties can be assigned deterministically. The $A_{\lambda}x_{\lambda}+B_{c}x_{c}$ terms supply this spin- and concentration-dependent component halo by halo, while $\mathcal{N}\!\left(0,\sigma_{\rm int}^{2}\right)$ is a stochastic representation of the remaining variation, which cannot be predicted from these two properties and the adopted linear model. By construction the two components reproduce the overall \HI\ scatter of the simulation; the Gaussian $\mathcal{N}\!\left(0,\sigma_{\rm int}^{2}\right)$ captures its width but not its detailed shape -- the mass dependence and the non-Gaussian, bimodal structure discussed next (Fig.~\ref{fig:scatter_c}). The fraction of the scatter carried by spin and concentration varies with halo mass and redshift, as we discuss below.

Figure~\ref{fig:scatter_c} compares the scatter of our fit with that of the simulation, at $z=0$ (upper) and $z=3.1$ (lower). The gray band is the 16--84 percentile spread of \HI\ measured directly in the simulation at fixed halo mass; the colored band is the spread produced by the fit (Eq.~\ref{eq:apply_c}) -- the median relation plus the spin and concentration terms and the intrinsic scatter. The two bands track each other closely, showing that the fit reproduces the amount of scatter in the simulation. At $z=0$ the scatter is largest near $M_{\rm h}\simeq10^{12}\,M_{\odot}$, the halo mass at which central galaxies are being quenched: at a given mass some centrals still hold their gas while others have already lost it, so the \HI\ values split into a gas-rich and a gas-poor group (a bimodal distribution) that broadens the scatter. The fit reproduces the overall width of this feature but not its asymmetry. The symmetric Gaussian cannot capture the extended gas-poor tail, where the gray band falls below the colored one; correspondingly, it produces the expected excess on the gas-rich side, where the colored band rises above the gray. This mismatch reflects the skewed, bimodal distribution near the quenching mass rather than the overall scatter amplitude.

The dependence on spin is positive at every redshift ($A_{\lambda}\simeq0.22$--$0.26$): at fixed halo mass, faster-spinning halos host more extended discs and hold on to more \HI. Spin is the most important secondary property along the gas-rich rising part of the relation, where it reduces the variance by a fraction $r^{2}_{\lambda}\simeq0.29$ at $z=0$, and $\simeq0.59$ by $z=5$ (at $M_{\rm h}\simeq10^{11.9}\,M_{\odot}$). Concentration acts in the opposite direction ($B_{c}<0$ at all redshifts): more concentrated, earlier-forming halos tend to be gas-poorer, and at $z=0$ concentration takes over as the leading secondary property at the massive end (it reduces the variance by up to $r^{2}_{c}\simeq0.16$ at $M_{\rm h}\simeq10^{12.9}\,M_{\odot}$). By $z\gtrsim3$ this concentration signal fades at the massive end ($r^{2}_{c}\lesssim0.02$), because halos are then only weakly concentrated ($c\simeq3$--$5$, near the minimum of the $V_{\rm max}/V_{\rm vir}$--$c$ relation, where the inversion is least sensitive). As a result the intrinsic scatter grows from $\sigma_{\rm int}\simeq0.31$~dex at $z=0$ to $\simeq0.36$~dex at $z=5$. This motivates replacing concentration by a direct formation-time proxy, which we explore in Appendix~\ref{sec:scatter_z50}. Where a catalog resolves substructure, the fit can be further extended to incorporate the satellite fraction (Appendix~\ref{sec:substructure_fraction}).

\begin{table}
\centering
\setlength{\tabcolsep}{4pt}
\caption{Best-fitting spin$+$concentration scatter model
(Eqs.~\ref{eq:plane_c}, \ref{eq:apply_c}) as a function of redshift.
$A_{\lambda}$ and $B_{c}$ are in dex per unit standardized predictor;
$\tilde\lambda, s_{\lambda}$ and $\tilde c, s_{c}$ are the median and spread
(half the 16--84 range) used to standardize spin and concentration;
$\sigma_{\rm int}$ is the scatter about the plane, in dex. Together with the
median relation (Table~\ref{tab:hihm}) these give the complete recipe of
Eq.~(\ref{eq:apply_c}). The fit uses the volume-weighted MSI$+$MSII sample
($N\simeq2$--$7\times10^{5}$ per redshift).}
\label{tab:scatter_c}
\begin{tabular}{lccccccc}
\hline\hline
$z$ & $A_{\lambda}$ & $B_{c}$ & $\tilde\lambda$ & $s_{\lambda}$ & $\tilde c$ & $s_{c}$ & $\sigma_{\rm int}$\\
\hline
0.0 & $\phantom{-}0.259$ & $-0.027$ & 0.037 & 0.023 & 11.12 & 3.46 & 0.311\\
1.0 & $\phantom{-}0.240$ & $-0.083$ & 0.035 & 0.020 & 7.65  & 2.80 & 0.319\\
2.1 & $\phantom{-}0.222$ & $-0.138$ & 0.038 & 0.021 & 4.93  & 2.44 & 0.331\\
3.1 & $\phantom{-}0.221$ & $-0.107$ & 0.040 & 0.022 & 3.57  & 1.69 & 0.343\\
3.9 & $\phantom{-}0.221$ & $-0.077$ & 0.041 & 0.022 & 2.93  & 1.35 & 0.353\\
4.9 & $\phantom{-}0.222$ & $-0.036$ & 0.041 & 0.022 & 2.51  & 1.17 & 0.361\\
\hline
\end{tabular}
\end{table}

\section{Summary and discussion}
\label{sec:conclusions}
We have used the GAEA2023 semi-analytic model \citep{delucia24b} to study the neutral-hydrogen content of galaxies and dark-matter halos across cosmic time. We have examined the \HI\ mass function (HIMF) and its decomposition by host-halo mass, the total \HI--halo mass relation $M_{\rm HI}(M_{\rm h})$ and its redshift evolution, its scatter, and the secondary halo properties associated with that scatter. Our aim has been to distil these results into a compact, physically motivated prescription for populating dark-matter-only halo catalogs with \HI\ for post-reionization 21-cm applications. Our main results are as follows:

\begin{enumerate}
    \item The GAEA2023 local HIMF reproduces the ALFALFA and HIPASS measurements above their completeness limits. Although the high-mass end was included among the model calibration constraints, the low-mass end, the weak evolution to $z\simeq1$, and the decomposition by host-halo mass were not. In particular, the latter agrees with the ALFALFA group measurements of \citet{jones20}, providing an independent validation of the predicted connection between \HI\ content and halo mass.

     \item The total \HI--halo mass relation at $z=0$ rises steeply, reaches a broad maximum near $M_{\rm h}\sim10^{11.7}\,M_\odot$, declines as central galaxies are quenched, and turns upward again toward group and cluster scales as satellites increasingly dominate the halo \HI\ budget. At high redshift, the relation approaches a nearly single power law, $M_{\rm HI}\propto M_{\rm h}$. A five-parameter analytic form reproduces the median relation to $\sim0.1$--$0.2$~dex at all redshifts considered.

     \item The scatter about the median is substantial ($\simeq0.5$~dex) and contains a systematic dependence on halo properties. At fixed halo mass, it correlates positively with the spin parameter $\lambda$ and negatively with the concentration $c$: spin dominates along the gas-rich rising branch, whereas concentration becomes more important at the massive end. A standardized linear plane in $\lambda$ and $c$, supplemented by a Gaussian residual with $\sigma_{\rm int}\simeq0.32$~dex at $z=0$, reproduces the overall variance of the \HI\ distribution.
\end{enumerate}

Direct observational constraints on $M_{\rm HI}(M_{\rm h})$ remain scarce and heterogeneous because the relation is inferred indirectly from group-based measurements, stacking analyses, and a limited number of individual systems, often using different halo-mass definitions. Nevertheless, our $z=0$ relation follows the bulk of the available estimates within their uncertainties (Fig.~\ref{fig:median_hihm_z0}). A complementary benchmark is provided by other galaxy-formation models, including GALFORM \citep{baugh19}, SHARK \citep{lagos18,chauhan20}, and the earlier GAEA2017 analysis of \citet{spinelli20}, with which we find qualitative and, over their common halo-mass range, quantitative agreement. We therefore regard the relation as well constrained where it is anchored by the HIMF and its halo-mass decomposition, and as a testable prediction elsewhere. Integrating it over halo mass predicts the cosmic \HI\ abundance, which can already be compared with damped-Ly$\alpha$ measurements of $\Omega_{\rm HI}$ at high redshift \citep{noterdaeme12,crighton15}. Other mass-weighted moments of the relation, together with its scatter, determine the \HI\ bias, shot noise, and clustering, providing complementary tests with forthcoming 21-cm observations.

Our analysis also clarifies the physical origin of the relation and its scatter. The characteristic shape of $M_{\rm HI}(M_{\rm h})$ reflects the successive importance of gas cooling, AGN-driven suppression of the central cold-gas reservoir, and the growing satellite contribution at high halo mass. At fixed mass, higher-spin and later-forming, less concentrated halos are systematically \HI-richer, consistent with angular-momentum support and extended gas-accretion histories favoring the retention of atomic gas. Earlier-forming halos and those hosting more massive black holes are instead \HI-poorer. The transition from a spin-dominated regime on the rising branch to an assembly- and AGN-dominated regime near and above the quenching scale, $M_{\rm h}\simeq10^{12},M_\odot$, is robust within GAEA2023 and qualitatively consistent with SHARK \citep{chauhan20}. This agreement across two different semi-analytic implementations supports the physical interpretation of the trends, although it is not a fully independent test because both models connect disc structure and angular momentum to the gas partition; their quantitative predictions also remain model-dependent.

One immediate application is to post-reionization 21-cm cosmology, whose large-scale signal depends on how \HI\ populates halos. Our median relation and its central--satellite decomposition provide the ingredients required by halo-model and halo-occupation prescriptions to assign \HI\ to dark-matter-only catalogs \citep{bagla10,padmanabhan17}. The scatter model additionally supplies a halo-property-dependent dispersion that is often neglected. Because its predictors are available in standard $N$-body catalogs, this component can be included without resolving galaxies or rerunning the galaxy-formation model. Accounting for its correlation with halo assembly should improve predictions for the clustering and shot noise of the \HI\ field, and hence the realism of mocks constructed for CHIME-, MeerKAT-, and SKA-class analyses \citep{bull15,SKA_SWG20}.

A more speculative application is to multi-messenger cosmology. The three-dimensional \HI\ density field measured through intensity mapping can provide a statistical redshift prior for gravitational-wave sources, as recently explored for binary black holes observed with the Einstein Telescope and \HI\ mapped by the SKA Observatory \citep{ulyana26}. Our prescription could inform the mock catalogs and forward models required for such analyses, although a quantitative application would require jointly modeling the gravitational-wave source population and the \HI\ field.

For precision cosmology, the value of such a prescription lies in controlling the astrophysical dependence of the 21-cm signal. Intensity-mapping measurements of baryon acoustic oscillations and redshift-space distortions probe the expansion history and growth rate, but their amplitudes constrain combinations involving $\Omega_{\rm HI}$ and $b_{\rm HI}$, while their interpretation also depends on scale-dependent bias and shot noise \citep{bull15,Villaescusa-Navarro:2018vsg}. Recent MeerKLASS results further demonstrate the importance of realistic mock-based validation, covariance estimation, and control of residual observational systematics \citep{meerklass25}. A physically grounded $M_{\rm HI}(M_{\rm h})$ relation with realistic scatter can help model these contributions and propagate their uncertainties into cosmological analyses. The same ingredients are relevant to proposed probes of the neutrino mass \citep{navarro15,pal16}, the nature of dark matter \citep{carucci15}, and primordial features \citep{xu16,ballardini18}. The present work nevertheless remains a single-cosmology, model-based prediction: it provides an astrophysical ingredient for such analyses rather than a cosmological constraint in itself.

Several limitations frame the future development of this program. Our results are based on a single WMAP1 cosmology. Although GAEA run with a Planck cosmology produces similar galaxy properties, both halo structure and the galaxy-formation response can depend on cosmological parameters; direct cosmological applications will therefore require recalibrating or emulating the relation across a range of cosmologies. The predictions are also model-dependent: despite the qualitative agreement with SHARK, the amplitudes of the median relation and its secondary dependences are sensitive to the adopted treatments of gas partitioning, environmental stripping, and AGN feedback. Finally, the linear-plane model with a Gaussian residual reproduces the overall variance efficiently but not the mass dependence, skewness, or bimodality of the residual distribution near the quenching scale. Addressing these limitations will be important as SKA precursors and, ultimately, the SKA Observatory move toward precision measurements of the post-reionization \HI\ distribution. The framework developed here provides a practical foundation for constructing the physically informed mock catalogs and controlling the astrophysical uncertainties on which that program will rely.

\begin{acknowledgements}
      MK acknowledges funding from INAF under the project ``Assegni Ricerca SKA CTA e Precursori'' (CUP: C54I19001050001; Objective Function: 1.05.03.32.06), project ID: 325/2024.
\end{acknowledgements}

%
   \bibliographystyle{bibtex/aa} 
   \bibliography{reference} 

\begin{appendix}




\makeatletter
\let\theapsection\relax
\let\theapsubsection\relax
\let\theapsubsubsection\relax
\makeatother

\appendix
\nolinenumbers
\section{The HIMF and \HI\ halo mass function by galaxy color}
\label{sec:himf_hihmf_color}

\subsection{The HIMF by galaxy color at \texorpdfstring{$z=0$}{z=0}}
\label{sec:himf_color}

\begin{figure}
	\includegraphics[width=\columnwidth]{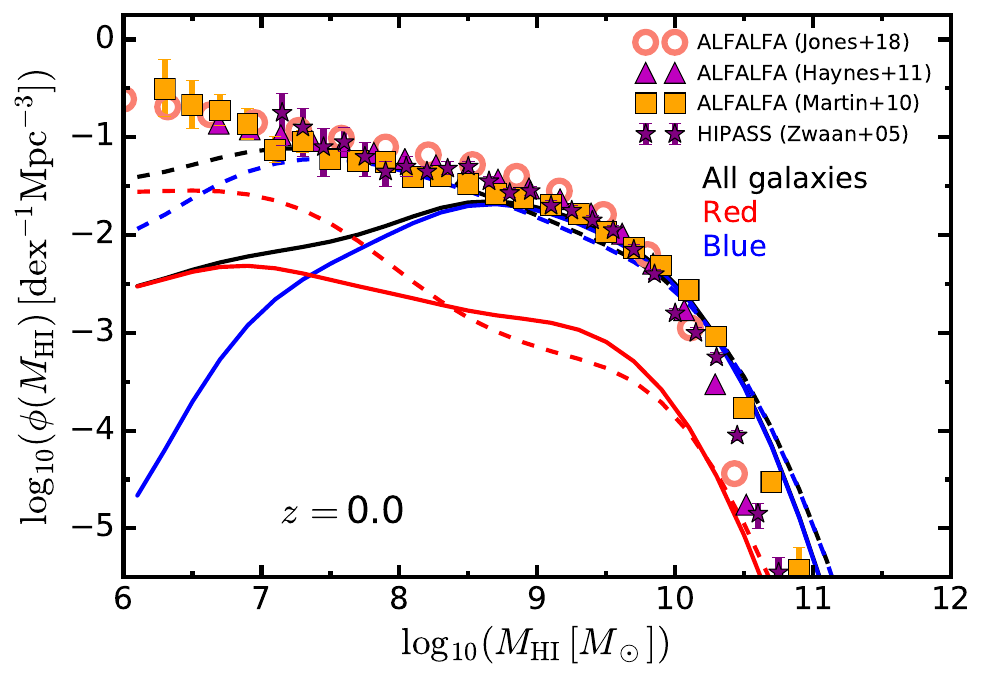}
    \caption{The galaxy HIMF at $z=0$ separated by galaxy color, including both central and satellite systems. Dashed black curves show the total HIMF (all galaxies), dashed red curves show red galaxies, and dashed blue curves show blue galaxies from the MSII simulation (equivalent MSI results, shown by solid lines, are nearly identical above the completeness limit). Observational data points from HIPASS \citep{zwaan05} and ALFALFA \citep{martin10,haynes11,jones18} are overplotted for the total population. Model predictions have been convolved with a representative $0.25$\,dex observational uncertainty in \HI-mass determination.}
    \label{fig:himf_allgal_redblue_z0}
\end{figure}

Figure~\ref{fig:himf_allgal_redblue_z0} shows the $z=0$ HIMF separated by galaxy color. Galaxies are classified as blue (star-forming) or red (quenched) by their specific star formation rate, adopting ${\rm sSFR}>0.3/t_{\rm H}$ for blue systems, with $t_{\rm H}$ the Hubble time \citep[see e.g.][]{gonzalez-perez17}. Blue galaxies dominate the total HIMF at nearly all \HI\ masses and therefore account for most of the cosmic \HI\ density. Red galaxies contribute comparably only at the most massive end ($M_{\rm HI}\gtrsim10^{10}\,M_\odot$), before declining more steeply toward lower masses, consistent with the reduced cold-gas reservoirs of early-type systems. The total HIMF reproduces the ALFALFA and HIPASS data above the survey completeness limit ($\log(M_{\rm HI}/M_\odot)\gtrsim9$); the observational points refer to the total population only, so the color decomposition is a pure model prediction. Below the completeness limit, the blue population continues smoothly down to $\log(M_{\rm HI}/M_\odot)\sim7$, whereas red galaxies become negligible, reflecting that low-\HI-mass systems are predominantly star-forming while quenched galaxies are largely confined to higher masses. The relative red and blue contributions should be interpreted with caution, as they depend on the adopted sSFR threshold and on the modeling of quenching, environmental stripping, and AGN feedback within the semi-analytic framework.

\subsection{The \HI--halo mass relation by galaxy color at \texorpdfstring{$z=0$}{z=0}}
\label{sec:mhihm_color}

\begin{figure}
	\includegraphics[width=\columnwidth]{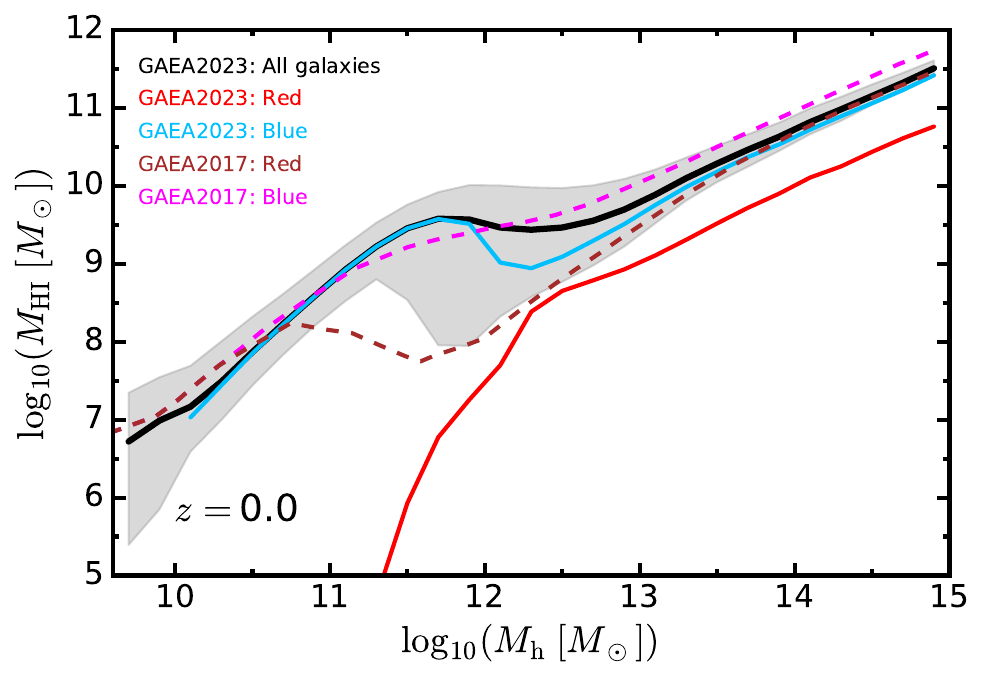}
    \caption{Median \HI--halo mass relation at $z=0$ for all galaxies and for red and blue populations. The color separation is as in Fig.~\ref{fig:himf_allgal_redblue_z0}. The black solid line shows the median $M_{\rm HI}(M_{\rm h})$ for all galaxies in GAEA2023. The red and deepskyblue solid lines indicate the corresponding relations for red and blue galaxies, respectively, while the brown and magenta dashed curves show the GAEA2017 predictions for red and blue galaxies as adopted by \citet{spinelli20}. The shaded region denotes the 16th--84th percentile range of $M_{\rm HI}$ for all galaxies in GAEA2023, obtained from MSII at $\log (M_{\rm h}/M_\odot)<11.9$ and from MSI at higher halo masses.}
    \label{fig:median_hihm_allgal_redblue_z0}
\end{figure}

Fig.~\ref{fig:median_hihm_allgal_redblue_z0} shows the median \HI--halo mass relation at $z=0$ for all galaxies in GAEA2023, decomposed into red and blue populations, and compares the predictions with those of the earlier GAEA2017 implementation of \citet{spinelli20}. In GAEA2023, the median \HI\ mass of all galaxies rises steeply with halo mass, reaches a local maximum of $\log(M_{\rm HI}/M_\odot)\simeq 9.6$ near $\log(M_{\rm h}/M_\odot)\simeq 11.7$, declines slightly toward $\log(M_{\rm h}/M_\odot)\simeq 12.3$, and then increases again toward higher masses; as discussed in Sect.~\ref{sec:Mh_MHI_diffz}, this renewed rise reflects the growing contribution of \HI-bearing satellite galaxies in massive halos. The blue population closely follows the all-galaxy relation at both low ($\log(M_{\rm h}/M_\odot)\lesssim 11.7$) and high ($\log(M_{\rm h}/M_\odot)\gtrsim 13$) halo masses, confirming that the halo \HI\ budget is dominated by star-forming systems in these regimes. Around the characteristic halo mass $\log(M_{\rm h}/M_\odot)\sim 12$, however, the blue relation exhibits a pronounced dip below the total before recovering at higher masses. This marks the halo-mass scale at which the dominant central galaxy progressively quenches and migrates from the blue to the red population; consequently, the median blue relation becomes increasingly dominated by star-forming satellite galaxies at higher masses, where it again approaches the total relation. By contrast, the red population remains \HI-poor over the entire halo-mass range: its median \HI\ mass declines rapidly below $\log(M_{\rm h}/M_\odot)\sim 11.5$ and increases gradually toward higher masses, but remains roughly $0.7$\,dex below the blue and total relations even at cluster scales. Red galaxies therefore make an appreciable contribution to the total halo \HI\ budget only in massive halos, without ever dominating it.

Relative to the GAEA2017 model of \citet{spinelli20}, GAEA2023 predicts slightly lower \HI\ masses for blue galaxies at $\log(M_{\rm h}/M_\odot)\gtrsim 12$, and the transition-mass dip seen in the blue population is absent from the earlier model. The two implementations differ most strongly for red galaxies: GAEA2017 predicts substantially more \HI\ in red systems at low and intermediate halo masses than GAEA2023, with the two converging only toward the highest halo masses. This likely reflects the revised treatment of environmental gas removal of \citet{xie20} -- gradual hot-gas stripping together with ram-pressure stripping of satellite cold gas -- present in GAEA2023 but absent from the \citet{xie17}-based configuration used by \citet{spinelli20}; \citet{xie18} showed that the earlier strangulation-only treatment depletes satellite \HI\ too inefficiently while over-depleting the molecular gas, which motivated the revision. We caution, however, that the \HI\ content of red galaxies at low halo masses is particularly sensitive to numerical resolution and to the modeling of orphan satellites, a long-standing GAEA ingredient \citep{delucia07} common to both model versions; below the MSII--MSI stitching scale, $\log(M_{\rm h}/M_\odot)\simeq 11.9$ (Sect.~\ref{sec:Mh_MHI_diffz}), the detailed shape of the red-galaxy relation should therefore be regarded as model-dependent rather than a robust physical prediction, and we refrain from drawing firm physical conclusions in this regime. The halo-to-halo variation in \HI\ content at fixed halo mass increases toward the low-mass end, where it likely reflects genuine diversity in gas-accretion and star-formation histories; we note, however, that this regime is sampled mainly by MSII near its resolution limit, so a numerical contribution to the scatter cannot be excluded.

\section{Interpreting the \texorpdfstring{\HI}{HI}--halo mass relation in GAEA2023--Cont.}
\label{sec:physical drivers cont}

\subsection{The halo spin parameter (\texorpdfstring{$\lambda_{\rm h}$)}{lambda}}
\label{sec:spin parameter cont}

\begin{figure}
    \centering
    \includegraphics[width=\columnwidth]{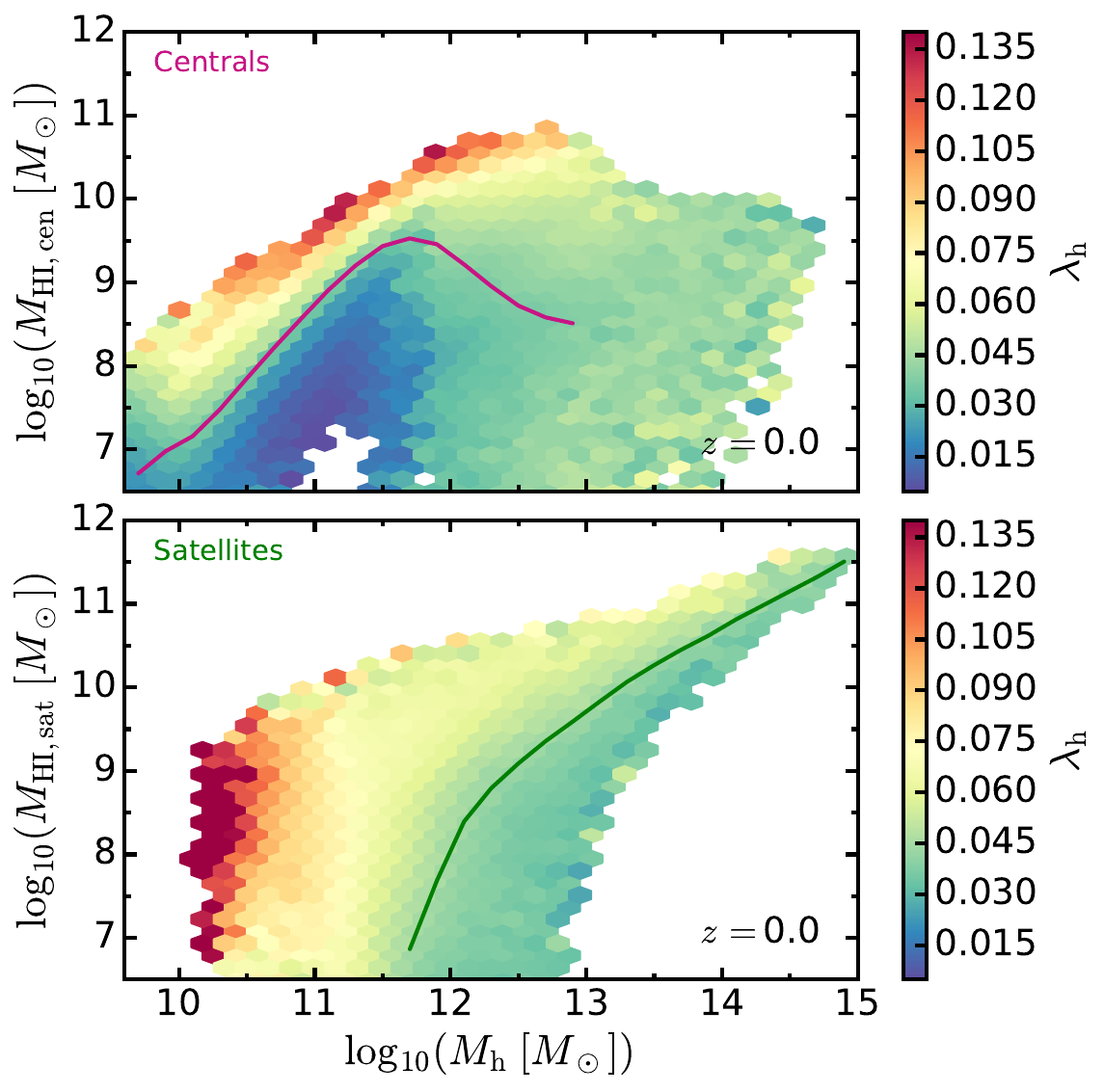}
    \caption{Same as Fig.~\ref{fig:hi_vs_CentralMvir_vs_spin}, but at $z=0$, and split by galaxy type: the \HI\ mass is restricted to central galaxies (\textit{upper}) and to satellite galaxies (\textit{lower}).}
    \label{fig:hi_vs_CentralMvir_vs_spin_cont}
\end{figure}

The upper panel of Fig.~\ref{fig:hi_vs_CentralMvir_vs_spin_cont} isolates the \HI\ content of central galaxies, with the magenta line showing their median $M_{\rm HI}(M_{\rm h})$ relation. In this projection the dependence on $\lambda_{\rm h}$ is more clearly visible than in the total relation of Sect.~\ref{sec:spin parameter}: at fixed $M_{\rm h}$, high-spin halos preferentially host \HI-rich centrals, whereas low-spin halos host \HI-poor ones. The separation is most pronounced around the halo-mass scale at which the central \HI\ mass peaks and begins to decline ($\log(M_{\rm h}/M_\odot)\simeq 11.7$), indicating that halo spin remains an important secondary parameter up to the onset of efficient quenching. This is consistent with a scenario in which higher-spin halos host more extended, lower-surface-density gas discs that favor the retention of atomic gas, although -- as cautioned in Sect.~\ref{sec:spin parameter} -- this correlation is partly built into the model through its angular-momentum-based disc prescription. At higher halo masses, where centrals are strongly quenched in GAEA2023, the dependence of central \HI\ on $\lambda_{\rm h}$ weakens markedly, consistent with a regime in which AGN feedback and the suppression of gas cooling dominate over angular-momentum support in regulating the cold-gas reservoir. This behavior is broadly consistent with the interpretation of \citet{chauhan20}, who likewise identified halo spin as an important secondary parameter for \HI-rich galaxies. 

The lower panel focuses on the satellite population, with the green line indicating the median $M_{\rm HI}(M_{\rm h})$ contributed by satellites in each halo. In contrast to the centrals, the dependence on $\lambda_{\rm h}$ at fixed halo mass is weak: across most of the $M_{\rm h}$--$M_{{\rm HI,sat}}$ plane, halos with different satellite \HI\ masses share similar median spin. This suggests that, in GAEA2023, the \HI\ content of satellites is only weakly coupled to the angular momentum of the host halo, and is instead regulated primarily by subhalo accretion histories and by environmental processes such as ram-pressure and tidal stripping acting on the gas after infall. Consequently, once the total \HI\ budget becomes satellite-dominated, the connection between halo spin and \HI\ content is substantially diluted, consistent with the weakening spin dependence seen at high halo mass in the total relation of Sect.~\ref{sec:spin parameter}. We caution that this panel is constructed from the MSI run alone, whose larger volume is needed to sample satellite-rich halos but which under-resolves low-mass systems; the apparent rise in median spin toward the lowest halo masses ($\log(M_{\rm h}/M_\odot)\lesssim 11$) is therefore dominated by a small number of poorly resolved halos and should not be interpreted as a physical trend. A comparably weak spin dependence for satellites would be expected if, as in the SHARK model \citep{chauhan20}, environment-driven processes are the leading regulators of satellite gas reservoirs.

\subsection{The halo formation history}
\label{sec:halo formation history}

Another halo property that may contribute to the scatter in the $M_{\rm HI}(M_{\rm h})$ relation is the halo formation time, which we characterize through $z_{50}$: the redshift at which the main progenitor first assembled half of the halo's mass at the epoch under consideration. Larger $z_{50}$ thus corresponds to earlier-forming halos, and smaller $z_{50}$ to later-forming, younger systems. Halo formation history is of particular interest in the context of assembly bias: from the clustering of \HI-selected ALFALFA galaxies, \citet{guo17} found that \HI-rich systems preferentially occupy later-forming halos at fixed halo mass, which they reproduced by adding formation time as a secondary halo-model parameter. Semi-analytic studies have likewise indicated that assembly history contributes to the scatter in the \HI--halo connection, although the strength of the trend and its mass dependence are model-dependent \citep{spinelli20, chauhan20}.

\begin{figure}
    \centering
    \includegraphics[width=\columnwidth]{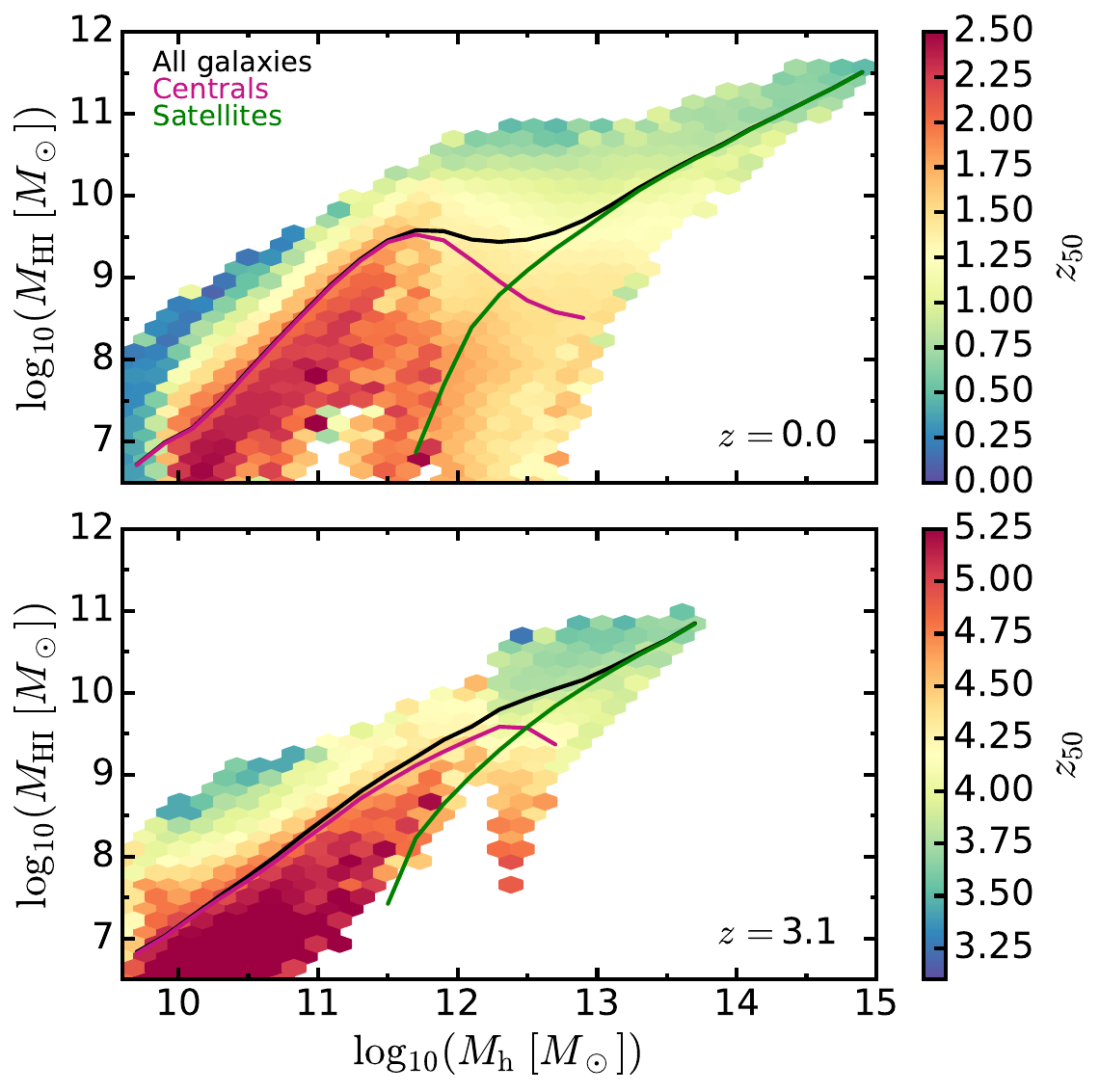}
    \caption{Same as Fig.~\ref{fig:hi_vs_CentralMvir_vs_spin}, but for the halo formation redshift $z_{50}$; larger $z_{50}$ indicates earlier-forming halos.}
    \label{fig:hi_vs_CentralMvir_vs_halo_formation_history}
\end{figure}

Fig.~\ref{fig:hi_vs_CentralMvir_vs_halo_formation_history} shows the distribution of $z_{50}$ in the $M_{\rm h}$--$M_{\rm HI}$ plane for the total \HI\ content of each parent halo, at $z=0$ (upper) and $z=3.1$ (lower). At $z=0$ a clear dependence on formation history is already present in the low-mass regime: at fixed halo mass, \HI-rich halos preferentially have smaller $z_{50}$ and are therefore typically younger than their \HI-poorer counterparts. The trend persists through the intermediate-mass (transition) regime but weakens markedly at the highest halo masses ($\log_{10}(M_{\rm h}/M_\odot)\gtrsim 13.5$). The same sense of the trend is recovered at $z=3.1$ -- where the whole population is shifted to higher $z_{50}$ because halos assemble earlier at high redshift -- indicating that the assembly dependence of the \HI\ content is not specific to the present epoch. Overall, these results suggest that halo formation history contributes to the scatter in halo \HI\ content primarily at low and intermediate halo masses, while its role becomes subdominant at the highest masses, where the total \HI\ reservoir is increasingly set by the cumulative satellite population and by stochastic environmental processing that washes out the imprint of halo assembly time.

\begin{figure}
    \centering
    \includegraphics[width=\columnwidth]{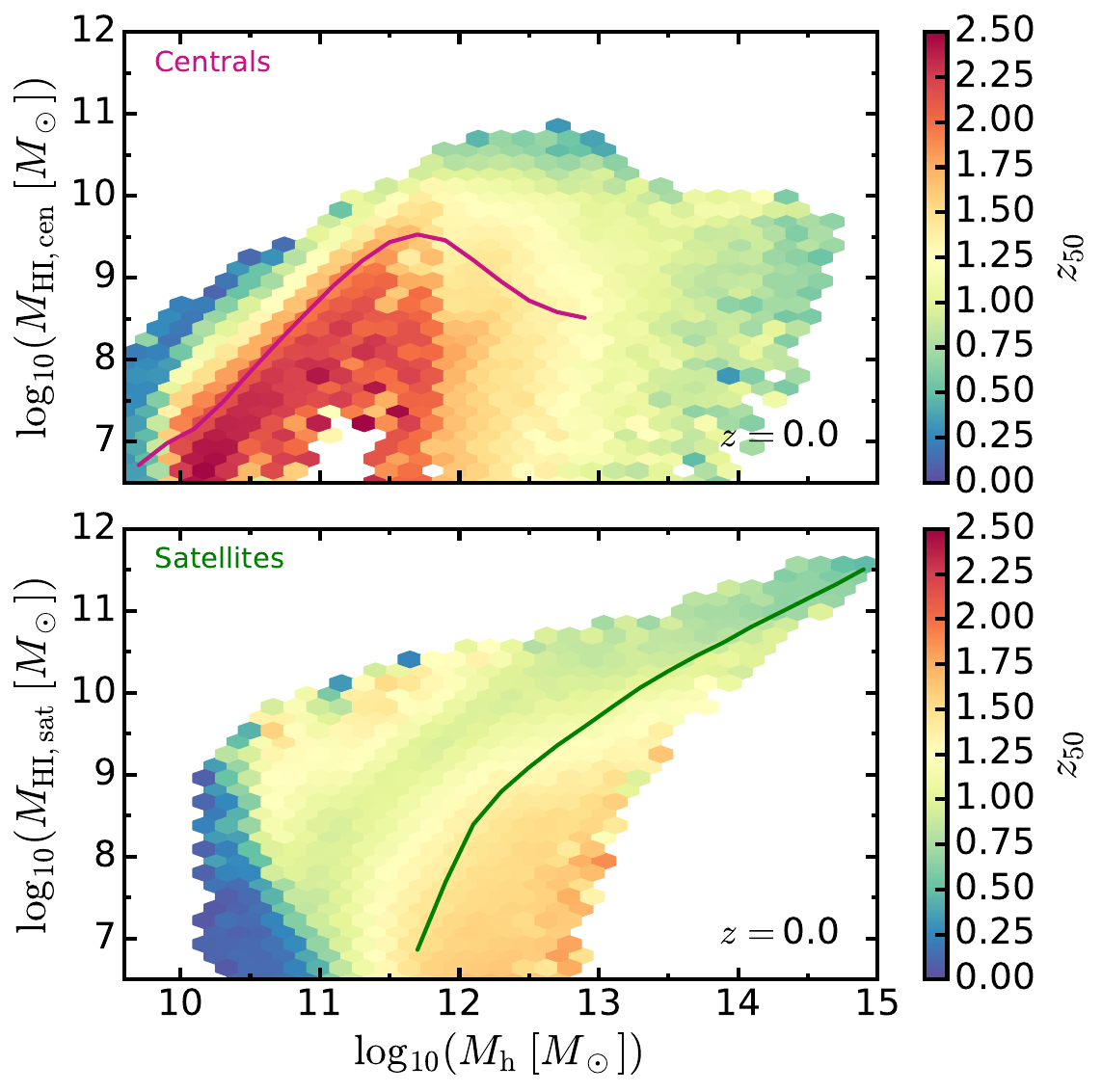}
    \caption{Same as Fig.~\ref{fig:hi_vs_CentralMvir_vs_spin_cont}, but colored by the formation redshift $z_{50}$ at $z=0$.}
    \label{fig:hi_vs_CentralMvir_vs_halo_formation_history_cent_sat}
\end{figure}

The upper panel of Fig.~\ref{fig:hi_vs_CentralMvir_vs_halo_formation_history_cent_sat} isolates the \HI\ content of central galaxies, with the magenta line showing their median $M_{\rm HI}(M_{\rm h})$ relation. The dependence on $z_{50}$ is strongest in the low-mass regime and remains visible up to $\log_{10}(M_{\rm h}/M_\odot)\sim 12$: at fixed halo mass, \HI-richer centrals are preferentially hosted by later-forming halos (smaller $z_{50}$). This indicates that, below the transition scale, the assembly history of the host halo is closely connected to the atomic-gas reservoir of its central galaxy, plausibly because later-forming halos continue to accrete gas, and consume it less rapidly, over a longer interval, allowing their centrals to retain more \HI\ at fixed $M_{\rm h}$. Above $\log_{10}(M_{\rm h}/M_\odot)\sim 12$ the dependence weakens substantially, consistent with the declining importance of the central \HI\ reservoir once the system enters the transition regime.

The lower panel focuses on the satellite contribution, with the green line showing the median satellite $M_{\rm HI}(M_{\rm h})$ relation. In contrast to the centrals, the dependence on formation history is weak at low halo masses but becomes clearly visible across the transition regime, $12\lesssim \log_{10}(M_{\rm h}/M_\odot)\lesssim 13.5$: there, at fixed halo mass, satellite systems with larger total \HI\ masses tend to occupy later-forming halos (smaller $z_{50}$). This suggests that, once the total halo \HI\ budget is no longer dominated by the central galaxy, the assembly dependence is carried primarily by the satellite population -- plausibly because satellites in later-forming halos have, on average, experienced environmental processing for a shorter time after infall, while those in earlier-forming halos have had longer to undergo stripping, starvation, and tidal evolution. At the highest halo masses the dependence again weakens, indicating that formation history is no longer a strong discriminator of the total satellite \HI\ reservoir. 

Taken together, these results indicate that the dependence of halo \HI\ content on formation history varies systematically across the halo-mass range: in the low-mass regime it is driven primarily by central galaxies, in the transition regime predominantly by satellites, and at the highest masses it becomes weak. Halo formation history is therefore an important secondary parameter shaping the scatter of the $M_{\rm HI}(M_{\rm h})$ relation below $\log_{10}(M_{\rm h}/M_\odot)\sim 13.5$, but not the dominant driver at the highest halo masses. This behavior is qualitatively consistent with earlier work linking the \HI\ content of galaxies and halos to assembly history \citep{guo17, spinelli20, chauhan20}; in particular, the central-dominated assembly dependence found here echoes the clustering-based result of \citet{guo17} that \HI-rich systems preferentially occupy later-forming halos. Finally, the trends discussed here should be interpreted as correlations rather than as evidence that halo formation history alone determines the \HI\ content of halos, since $z_{50}$, halo concentration, and halo spin are themselves strongly inter-correlated. As for concentration (Sect.~\ref{sec:halo concentration}), $z_{50}$ is not an explicit ingredient of the GAEA gas model, so its correlation with \HI\ content emerges indirectly through halo assembly.

\subsection{The black-hole to stellar-mass ratio}
\label{sec:bh_to_stellar}

A further galaxy property that may help explain the shape and scatter of the \HI--halo mass relation is the ratio between the central black-hole mass and the stellar mass of the central galaxy, $M_{\rm BH}/M_\ast$. This ratio provides a proxy for the cumulative growth of the black hole relative to the assembly of its host galaxy, and hence for the long-term impact of AGN feedback on the cold-gas reservoir. In GAEA2023, AGN feedback acts in two modes \citep{fontanot20}: a radio (hot-halo) mode, in which accretion onto the central black hole offsets the cooling flow in massive halos and curbs the resupply of cold gas, at a rate set by $M_{\rm BH}$ and the halo virial properties (their Eq.~2); and a quasar mode, in which cold-gas accretion drives outflows that expel cold gas. Both deplete the cold-gas -- and hence \HI\ -- reservoirs of massive centrals, and the radio mode may dominate at low redshift, where the high-accretion events powering the quasar mode grow rarer. We emphasize that $M_{\rm BH}/M_\ast$ is not itself an input to these prescriptions but serves as an empirical tracer of their cumulative effect. Larger values of $M_{\rm BH}/M_\ast$ are therefore expected to be associated with reduced \HI\ content, particularly in the halo-mass regime where AGN feedback becomes effective.

\begin{figure}
    \centering
    \includegraphics[width=\columnwidth]{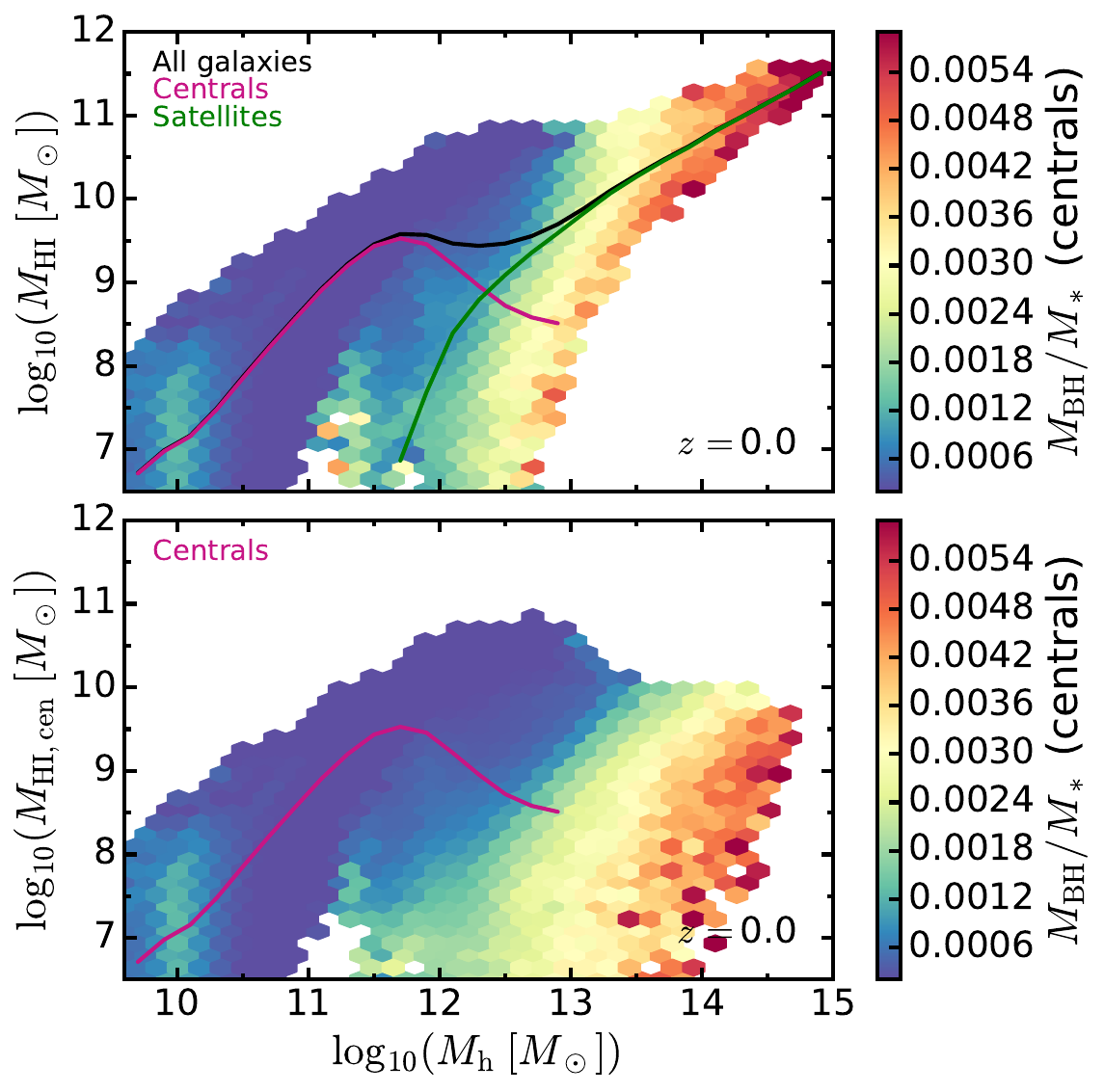}
    \caption{Same as Fig.~\ref{fig:hi_vs_CentralMvir_vs_spin}, but colored by the central black-hole to stellar-mass ratio $M_{\rm BH}/M_\ast$ at $z=0$: the binned \HI\ mass is the total halo content (\textit{upper}) and that of the central galaxy only (\textit{lower}).}
    \label{fig:hi_vs_mbh_to_mstar}
\end{figure}

Figure~\ref{fig:hi_vs_mbh_to_mstar} shows the median $M_{\rm BH}/M_\ast$ of central galaxies in the $M_{\rm h}$--$M_{\rm HI}$ plane at $z=0$. Both panels show the same qualitative trend: at fixed halo mass, systems with larger $M_{\rm BH}/M_\ast$ tend to be more \HI-poor. The dependence is weak at low halo masses, strengthens across the intermediate-mass (transition) regime, and remains visible for the central \HI\ content toward higher masses. It is clearest in the lower panel: once $\log_{10}(M_{\rm h}/M_\odot)\gtrsim 12$, \HI-poor centrals are systematically associated with larger $M_{\rm BH}/M_\ast$. This is consistent with a picture in which black-hole growth, together with the associated AGN-driven outflows, increasingly depletes the cold-gas reservoir of the central galaxy -- and hence its atomic component -- through the transition regime. We caution, however, that at low halo masses the central black-hole masses are small and largely set by the seeding and early-growth prescription, so $M_{\rm BH}/M_\ast$ is a less informative discriminator in this regime and the weak low-mass trend should not be over-interpreted.

The same tendency is present for the total halo \HI\ budget (upper panel), but is less sharply defined than in the central-only projection. This is expected, since $M_{\rm BH}/M_\ast$ refers only to the central galaxy while the total \HI\ mass also includes satellites. At low halo masses, where the total \HI\ content remains dominated by the central, the two projections largely agree; across the transition regime the anti-correlation is progressively broadened by the growing satellite contribution. At the highest masses the total \HI\ content rises with $M_{\rm h}$ because it is dominated by satellites, while the central $M_{\rm BH}/M_\ast$ also increases with halo mass; the residual dependence at fixed halo mass therefore weakens, and the link between the halo-integrated \HI\ content and the central black-hole properties becomes increasingly indirect. We note that only the black hole hosted by the central galaxy is considered here, since its accretion-driven outflows dominate the AGN feedback affecting the central galaxy's cold-gas reservoir in the GAEA framework.

These results are qualitatively consistent with \citet{chauhan20}, who identified $M_{\rm BH}/M_\ast$ as a secondary parameter closely associated with the scatter of the \HI--halo mass relation in the transition regime of the SHARK model. Their maps show similar behavior: in the halo-integrated projection, the anti-correlation at fixed halo mass is strongest through the transition regime and gives way, at the highest masses, to a satellite-driven upturn along which the central $M_{\rm BH}/M_\ast$ is uniformly high; in the central-galaxy projection, expressed against subhalo mass, the anti-correlation strengthens through the transition regime and persists toward higher masses. As in GAEA2023, the dependence is expressed far more clearly for centrals than for the halo-integrated \HI\ budget, consistent with AGN feedback acting directly on the central cold-gas reservoir around and above the characteristic mass scale where the median central \HI\ content peaks and begins to decline.

Overall, Fig.~\ref{fig:hi_vs_mbh_to_mstar} suggests that the black-hole to stellar-mass ratio is a useful diagnostic of the processes shaping the scatter of the \HI--halo mass relation, tracing the increasing importance of AGN feedback along the central-galaxy branch as the median central \HI\ content declines through the transition regime. It is not, however, adopted as a predictor in our scatter model: the prescription presented in Sect.~\ref{sec:hihm_scatter_fit} is deliberately restricted to quantities available directly from a dark-matter halo catalog -- such as halo spin, formation time, and concentration -- so that it remains applicable to analyses that populate dark-matter halos with \HI\ (for example, to generate 21-cm mocks) without requiring detailed galaxy-formation output. As for halo spin and concentration, the trends should be read as correlations rather than as evidence for a unique causal role of $M_{\rm BH}/M_\ast$, since black-hole growth, stellar mass, halo mass, and assembly history are mutually correlated. The weaker and more diffuse dependence in the halo-integrated projection further indicates that, on its own, $M_{\rm BH}/M_\ast$ cannot account for the full halo \HI\ budget once satellite galaxies become the dominant contributors.

\subsection{The substructure mass fraction}
\label{sec:substructure_fraction}

A further quantity that may contribute to the scatter of the \HI--halo mass relation is the ratio between the total mass in satellite subhalos and the parent halo mass, $M_{\rm h}^{\rm sat}/M_{\rm h}$. It therefore provides a useful tracer of the extent to which the halo \HI\ budget may be supplemented by its satellite population. Since satellites contribute an increasing fraction of the total \HI\ content toward high halo masses, this quantity is expected to become most relevant once the halo-integrated \HI\ reservoir is no longer dominated by the central galaxy. As with the black-hole to stellar-mass ratio, we examine $M_{\rm h}^{\rm sat}/M_{\rm h}$ to interpret the physical origin of the scatter rather than as an ingredient of our scatter prescription (Sect.~\ref{sec:hihm_scatter_fit}), since it requires resolved subhalo information unavailable in many dark-matter halo catalogs.

\begin{figure}
    \centering
    \includegraphics[width=\columnwidth]{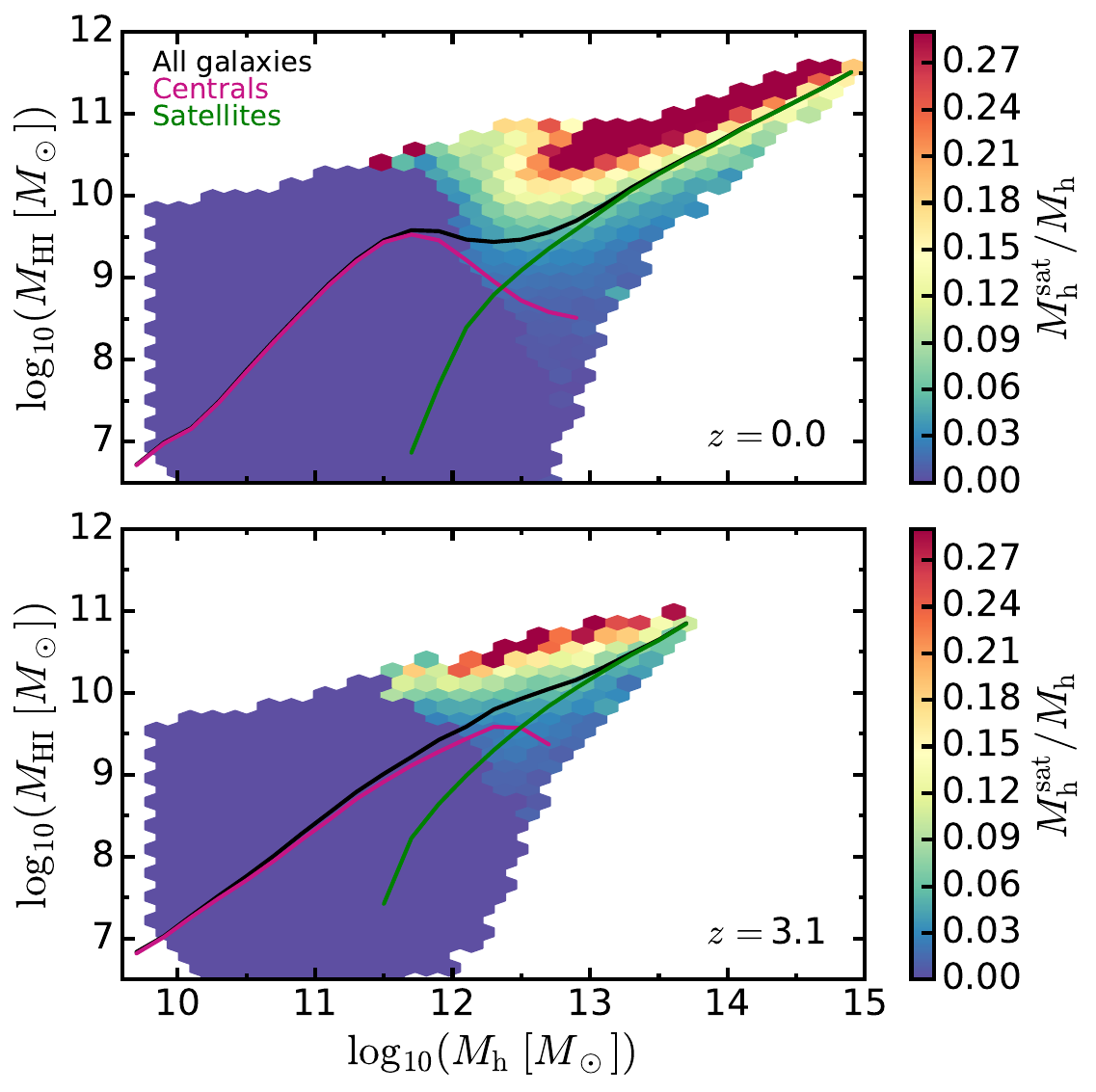}
    \caption{Same as Fig.~\ref{fig:hi_vs_CentralMvir_vs_spin}, but for the substructure mass fraction $M_{\rm h}^{\rm sat}/M_{\rm h}$, the total mass in satellite subhalos relative to the parent halo mass.}
    \label{fig:hi_vs_CentralMvir_vs_ratio_Msubh_to_Mh}
\end{figure}

Figure~\ref{fig:hi_vs_CentralMvir_vs_ratio_Msubh_to_Mh} shows the median $M_{\rm h}^{\rm sat}/M_{\rm h}$ in the $M_{\rm h}$--$M_{\rm HI}$ plane at $z=0$ (upper) and $z=3.1$ (lower). A clear mass dependence is apparent at both redshifts. In the low-mass regime the substructure mass fraction is close to zero over most of the populated region, indicating that these halos contain little mass in resolved satellite subhalos and that their \HI\ content is correspondingly dominated by the central galaxy; here $M_{\rm h}^{\rm sat}/M_{\rm h}$ contributes little to the scatter of the \HI--halo mass relation. We caution that the smallest subhalos fall below the resolution limit of the simulations, so the inferred substructure fraction at low halo masses should be regarded as a lower bound; the near-zero values are nonetheless consistent with the expectation that low-mass halos host relatively little bound substructure.

The behavior changes across the intermediate-mass (transition) regime. At fixed halo mass, systems with larger total \HI\ masses tend to exhibit larger values of $M_{\rm h}^{\rm sat}/M_{\rm h}$. The correlation becomes strongest once the satellite contribution begins to rise rapidly, near the halo-mass range where the median total \HI--halo mass relation flattens and the central-galaxy \HI\ reservoir starts to decline. This indicates that part of the scatter in the halo-integrated \HI\ content reflects differences in the surviving substructure population, which in turn traces the cumulative contribution of satellites to the total halo \HI\ reservoir.

In the high-mass regime the same qualitative behavior continues and is particularly evident along the upper envelope of the distribution: at fixed halo mass, \HI-richer halos systematically exhibit larger values of $M_{\rm h}^{\rm sat}/M_{\rm h}$. Physically, this is expected if halos retaining a larger fraction of their mass in surviving subhalos also host a larger reservoir of \HI-bearing satellites, thereby maintaining a higher halo-integrated \HI\ content even after the central galaxy has become strongly \HI-poor. In this sense, the substructure fraction does not directly regulate the \HI\ content of the central galaxy; rather, it traces the degree to which the total halo \HI\ budget is supplemented by satellites, supporting the interpretation that the upturn of the total \HI--halo mass relation at high halo masses is driven by the growing importance of the satellite population. The lower panel shows that this behavior is already established at $z=3.1$, indicating that the association between substructure fraction and halo \HI\ content is not confined to the present epoch.

These results are qualitatively consistent with those of \citet{chauhan20}, who identified the substructure mass fraction as one of the secondary parameters most closely associated with the scatter of the \HI--halo mass relation in the high-mass regime of the SHARK model, with larger values of $M_{\rm h}^{\rm sat}/M_{\rm h}$ corresponding to larger halo \HI\ masses over the redshift range $0\lesssim z\lesssim2$. Our results recover the same overall behavior and extend it to $z=3.1$. In GAEA2023, however, the correlation becomes clearly visible already through the transition regime, indicating that the growing importance of the satellite population begins to shape the scatter before halos become fully satellite-dominated. We emphasize that this dependence should be interpreted as an association rather than as evidence that substructure directly regulates the gas content of galaxies. Since the surviving substructure population is itself closely linked to halo assembly history, part of the correlation reported here likely reflects the same assembly-driven trends discussed in Appendix~\ref{sec:halo formation history}; the substructure mass fraction should therefore be regarded as a complementary diagnostic of the halo \HI\ budget rather than an independent physical driver of the scatter.

\section{\HI\ halo mass function: Fitting relations--Cont}
\label{sec:censat_fit}

\subsection{Fitting the satellite HI--halo mass relation}
\label{sec:satellite_hihm_fit}
With the aim of providing the ingredients needed by HOD techniques that simulate 21-cm maps for intensity mapping, we extend the analysis of Sect.~\ref{sec:hihm_fit} by modeling the \HI--halo mass relation of central and satellite galaxies separately.

\emph{Centrals.} We use the same functional form as for the total relation, Eq.~\ref{eq:hihm}, again fixing $\gamma=0.5$ and $M_{\rm min}=10^{8}\,M_\odot$ (the low-mass cut-off is unresolved for centrals), and stitching MSII (low mass) to MSI (high mass). The best-fitting parameters are listed in Table~\ref{tab:cen}. In contrast to the total relation, the central turnover is well defined at all redshifts, and $M_{\rm break}$ is robustly constrained: it \emph{increases} with redshift, from $\log_{10}(M_{\rm break}/M_\odot)\simeq11.4$ at $z=0$ to $\simeq12.5$ at $z\simeq5$, i.e.\ the halo mass at which central \HI\ is quenched shifts to higher values at earlier times. The high-mass amplitude $a_2$ is small and may be negative, reflecting the decline of central \HI\ toward massive halos (no satellite-driven upturn is present for centrals).

\emph{Satellites.} The satellite relation is well described by the simpler form of
\citet{spinelli20},
\begin{equation}
M_{\rm HI}^{\rm sat}(M_{\rm h}) = M_0 \left(\frac{M_{\rm h}}{M_{\rm min}}\right)^{\alpha} \exp\!\left[-\left(\frac{M_{\rm min}}{M_{\rm h}}\right)^{\gamma}\right] , 
\label{eq:hihm_sat}
\end{equation}
a power law of slope $\alpha$ normalized to $M_0$ at $M_{\rm h}=M_{\rm min}$, with a low-mass cut-off at $M_{\rm min}$ of sharpness $\gamma$ (here all four parameters are free). Because the satellite-dominated regime lies at high halo mass, we fit the larger MSI box only, which provides the statistics needed for the rare, satellite-rich halos. The best-fitting values are given in Table~\ref{tab:sat}. The slope is close to linear ($\alpha\simeq0.85$--$1.0$ at all $z$), and the cut-off mass $M_{\rm min}\sim10^{12}\,M_\odot$ marks the host-halo scale below which halos contain too few satellites to contribute appreciable \HI.

\begin{figure}
\centering
\includegraphics[width=\columnwidth]{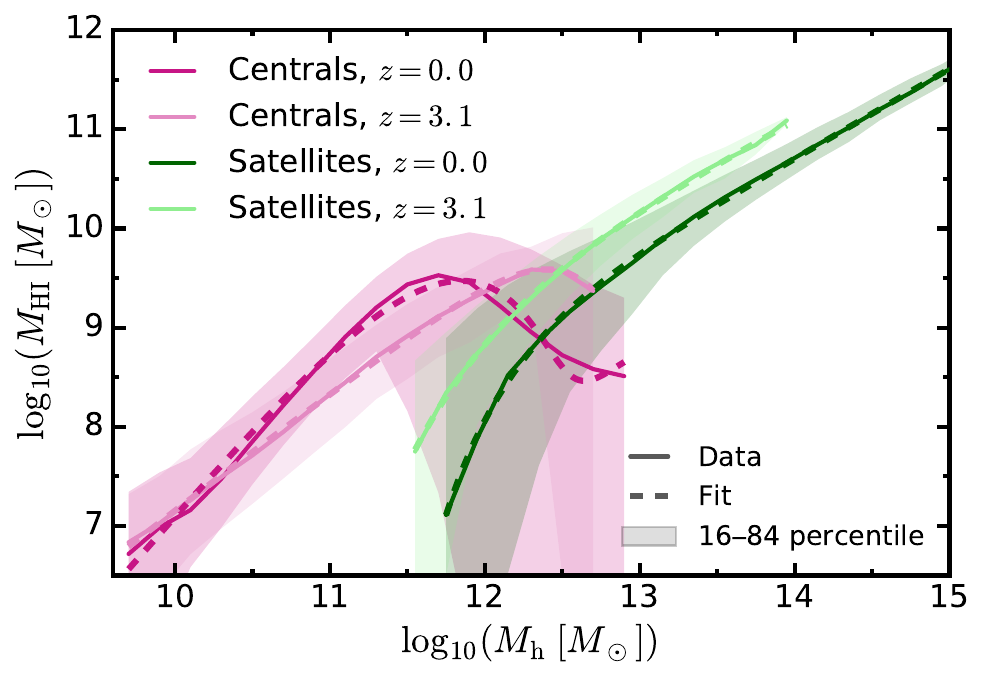}
\caption{Central (magenta) and satellite (green) \HI--halo mass relations for GAEA2023 at $z=0$ (dark shade) and $z=3.1$ (light shade). Solid lines: GAEA median; shaded bands: 16--84th percentile scatter; dashed lines: best-fitting Eq.~\ref{eq:hihm} (centrals, Table~\ref{tab:cen}) and Eq.~\ref{eq:hihm_sat} (satellites, Table~\ref{tab:sat}). 
}
\label{fig:censat}
\end{figure}

Figure~\ref{fig:censat} compares the GAEA medians (solid, with 16--84th percentile scatter shaded) with the fits (dashed) for centrals and satellites at $z=0$ and $z=3.1$.

\begin{table}
\centering
\setlength{\tabcolsep}{4pt}
\caption{Best-fitting parameters of the \emph{central} \HI--halo mass relation, Eq.~\ref{eq:hihm}, at the GAEA2023 snapshots closest to $z=0$--$5$. Masses in physical $M_\odot$ ($h=0.73$); $a_1,a_2$ dimensionless. $M_{\rm min}=10^{8}\, M_\odot$ and $\gamma=0.5$ are held fixed.}
\label{tab:cen}
\begin{tabular}{lccccc}
\hline\hline
$z$ & $a_1$ & $a_2$ & $\alpha$ & $\beta$ & $\log_{10}(M_{\rm break}/M_\odot)$ \\
\hline
0.0 & $1.8\times10^{-3}$ & $5.7\times10^{-5}$  & 0.80 & 0.79 & 11.39 \\
1.0 & $1.5\times10^{-3}$ & $1.3\times10^{-5}$  & 1.34 & 0.42 & 12.15 \\
2.1 & $2.0\times10^{-3}$ & $-2.4\times10^{-4}$ & 1.74 & 0.18 & 12.46 \\
3.1 & $1.6\times10^{-3}$ & $6.5\times10^{-5}$  & 1.40 & 0.15 & 12.45 \\
3.9 & $2.4\times10^{-3}$ & $-8.7\times10^{-4}$ & 0.72 & 0.13 & 12.49 \\
4.9 & $4.9\times10^{-3}$ & $-2.8\times10^{-3}$ & 0.35 & 0.11 & 12.53 \\
\hline
\end{tabular}
\end{table}

\begin{table}
\centering
\caption{Best-fitting parameters of the \emph{satellite} \HI--halo mass relation,
Eq.~\ref{eq:hihm_sat} (MSI only), at the GAEA2023 snapshots closest to
$z=0$--$5$. Masses in physical $M_\odot$ ($h=0.73$).}
\label{tab:sat}
\begin{tabular}{lcccc}
\hline\hline
$z$ & $\log_{10}(M_0/M_\odot)$ & $\log_{10}(M_{\rm min}/M_\odot)$ & $\alpha$ & $\gamma$ \\
\hline
0.0 & 9.15 & 12.27 & 0.90 & 1.08 \\
1.0 & 9.27 & 12.00 & 0.87 & 1.23 \\
2.1 & 9.46 & 12.09 & 0.87 & 0.89 \\
3.1 & 9.61 & 12.21 & 0.83 & 0.72 \\
3.9 & 8.98 & 11.86 & 1.02 & 1.02 \\
4.9 & 8.88 & 11.82 & 1.01 & 1.06 \\
\hline
\end{tabular}
\end{table}

\begin{figure}
\centering
\includegraphics[width=\columnwidth]{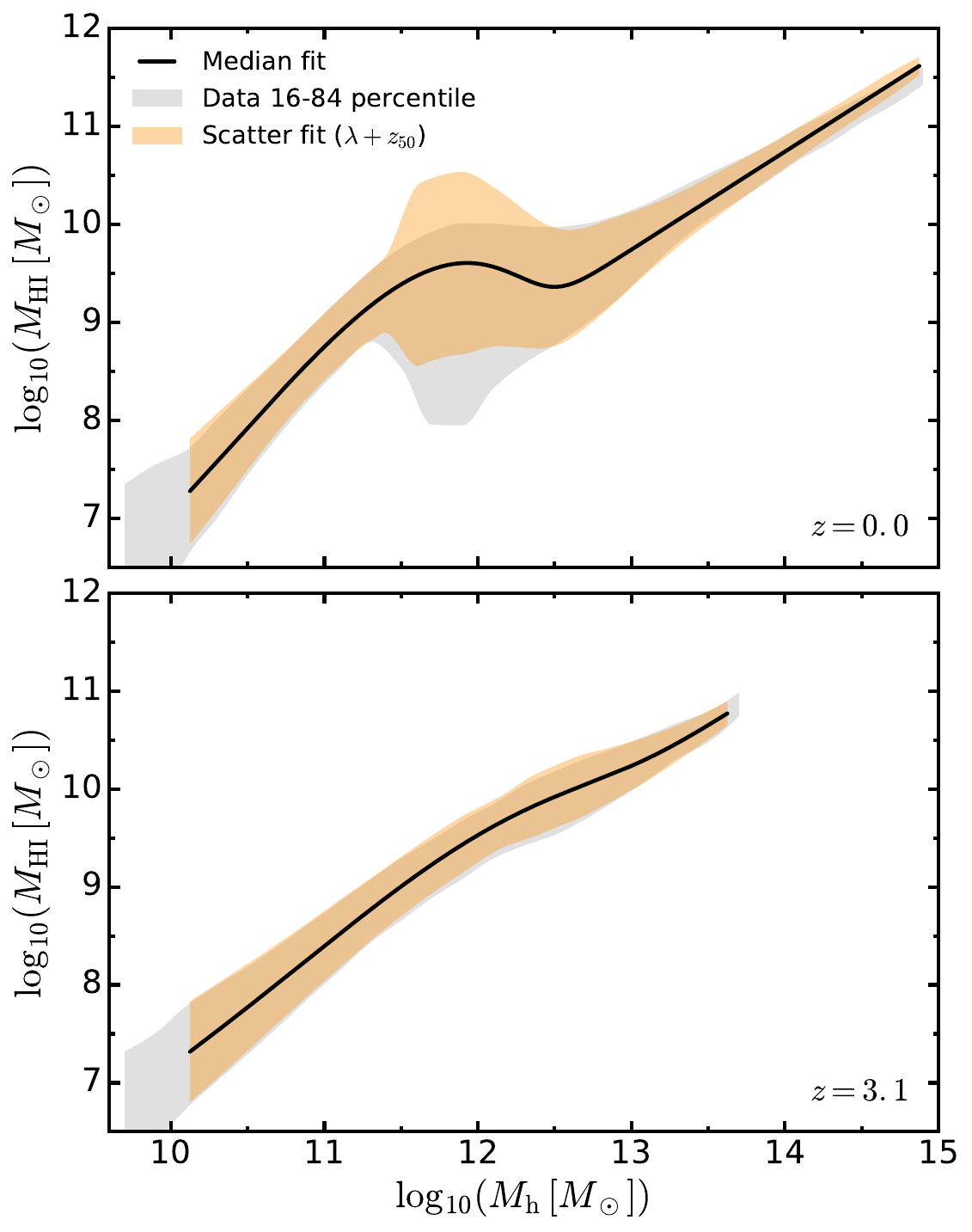}
\caption{Total \HI--halo mass relation and its scatter conditioned on spin and formation time at $z=0$ (upper) and $z=3.1$ (lower). Solid line: median relation; gray band: simulated $16$--$84$ scatter; orange band: total scatter reconstructed from the $\lambda+z_{50}$ plane.}
\label{fig:scatter_z50}
\end{figure}

\subsection{The scatter conditioned on spin and formation time}
\label{sec:scatter_z50}

As an alternative to concentration, we model the scatter of the total \HI--halo mass relation using the halo spin $\lambda$ and the formation redshift ($z_{50}$; Appendix~\ref{sec:halo formation history}), a direct measure of halo assembly taken from the merger trees. We follow the identical procedure of Sect.~\ref{sec:hihm_scatter_fit}: for the deviation $\Delta$ of Eq.~(\ref{eq:delta_c}) we fit, at each redshift, the plane

\begin{equation}
\Delta = C + A_{\lambda}\,x_{\lambda} + B_{z_{50}}\,x_{z_{50}},
\label{eq:plane_z50}
\end{equation}
with $x_{\lambda}$ and $x_{z_{50}}$ standardized as in Eq.~(\ref{eq:standardise_c}). Because $z_{50}$ can be undefined for poorly resolved assembly histories, the fit is restricted to halos with a well-defined formation time. The best-fitting coefficients and standardization constants are listed in Table~\ref{tab:scatter_z50}, and the relation is applied to a halo catalog exactly as in Eq.~(\ref{eq:apply_c}) with $x_{c}$ replaced by $x_{z_{50}}$. Figure~\ref{fig:scatter_z50} shows, at $z=0$ and $z=3.1$, the median relation with the simulated $16$--$84$ scatter (gray) and the total scatter reconstructed from the $\lambda+z_{50}$ model (median$\,\pm\,\sigma_{\rm total}$); the two agree, confirming that the plane reproduces the total scatter, apart from the same $z=0$ transition asymmetry seen in Fig.~\ref{fig:scatter_c}.

As for concentration, the spin response is positive at all redshifts ($A_{\lambda}\simeq0.20$--$0.25$) and dominates the gas-rich rising branch. Formation time acts with the opposite sign at every redshift ($B_{z_{50}}<0$): earlier-forming halos are systematically gas-poorer at fixed halo mass, and the effect strengthens monotonically with redshift, from $B_{z_{50}}=-0.08$ at $z=0$ to $-0.20$ at $z=5$. In contrast to concentration, formation time retains its discriminating power at high redshift, so that the intrinsic scatter conditioned on $\lambda+z_{50}$ is $\sigma_{\rm int}\simeq0.30$--$0.31$~dex and is essentially independent of redshift (Table~\ref{tab:scatter_z50}). Formation time and concentration probe the same assembly effect where both are informative---their variance reductions become comparable by $z\simeq1$---but $z_{50}$ remains a robust secondary out to $z=5$, whereas concentration does not. \citet{chauhan20} found $z_{50}$ less effective than spin and substructure and did not adopt it in their model; we nonetheless retain it as a single tree-based assembly proxy that remains informative out to $z=5$, beyond the $0\le z\le2$ range they considered.

\begin{table}
\centering
\caption{As Table~\ref{tab:scatter_c}, for the spin$+$formation-time plane (Eq.~\ref{eq:plane_z50}), with the formation-time constants $\tilde z_{50},\,s_{z_{50}}$ replacing the concentration constants $\tilde c,\,s_{c}$. The fit uses the halos with a well-defined $z_{50}$, $N=6.4,\,6.9,\,4.0,\,3.0,\,2.1,\,1.6\times10^{5}$ from $z=0$ to $5$.}
\label{tab:scatter_z50}
\begin{tabular}{lccccccc}
\hline\hline
$z$ & $A_{\lambda}$ & $B_{z_{50}}$ & $\tilde\lambda$ & $s_{\lambda}$ & $\tilde z_{50}$ & $s_{z_{50}}$ & $\sigma_{\rm int}$\\
\hline
0.0 & $\phantom{-}0.253$ & $-0.078$ & 0.037 & 0.023 & 1.87 & 0.94 & 0.300\\
1.0 & $\phantom{-}0.233$ & $-0.127$ & 0.035 & 0.020 & 2.72 & 0.88 & 0.307\\
2.1 & $\phantom{-}0.212$ & $-0.168$ & 0.038 & 0.021 & 3.81 & 0.89 & 0.311\\
3.1 & $\phantom{-}0.205$ & $-0.185$ & 0.040 & 0.022 & 4.81 & 0.88 & 0.313\\
3.9 & $\phantom{-}0.200$ & $-0.197$ & 0.041 & 0.022 & 5.64 & 0.86 & 0.312\\
4.9 & $\phantom{-}0.197$ & $-0.198$ & 0.041 & 0.022 & 6.70 & 0.80 & 0.314\\
\hline
\end{tabular}
\end{table}

\end{appendix}
\end{document}